\documentclass[%
 aip,
 amsmath,amssymb,
 reprint,%
]{revtex4-2}

\usepackage{graphicx}% Include figure files
\usepackage{dcolumn}% Align table columns on decimal point
\usepackage{bm}% bold math
\usepackage[utf8]{inputenc}
\usepackage[T1]{fontenc}
\usepackage{mathptmx}

\begin{document}
\preprint{AIP/123-QED}

\title[Scaling Theory of 3D Magnetic Reconnection Spreading]{Scaling Theory of Three-Dimensional Magnetic Reconnection Spreading}

\author{M. Arencibia}
\affiliation{Department of Physics and
  Astronomy, West Virginia University, Morgantown, WV 26506, USA}
\email{miarencibia@mix.wvu.edu}  

\author{P. A. Cassak} \affiliation{Department of Physics and
  Astronomy and Center for KINETIC Plasma Physics, West Virginia University, Morgantown, WV 26506, USA}
\email{Paul.Cassak@mail.wvu.edu}

\author{M. A. Shay}
\affiliation{Department of Physics and Astronomy, University of Delaware, Newark, Delaware 19716, USA}
 
  \author{E. R. Priest}
  \affiliation{St Andrews University
Mathematics Department
St Andrews University
St Andrews KY16 8QR
UK}

\begin{abstract}
We develop a first-principles scaling theory of the spreading of three-dimensional (3D) magnetic reconnection of finite extent in the out of plane direction. This theory addresses systems with or without an out of plane (guide) magnetic field, and with or without Hall physics. The theory reproduces known spreading speeds and directions with and without guide fields, unifying previous knowledge in a single theory. New results include: (1) Reconnection spreads in
a particular direction if an x-line is induced at the interface between reconnecting and non-reconnecting regions, which
is controlled by the out of plane gradient of the electric field in the outflow direction. (2) The spreading mechanism for anti-parallel collisionless reconnection is convection, as is known, but for guide field reconnection it is magnetic field bending. We confirm the theory using 3D two-fluid and resistive-magnetohydrodynamics simulations. (3) The theory explains why anti-parallel reconnection in resistive-magnetohydrodynamics does not spread. (4) The simulation domain aspect ratio, associated with the free magnetic energy, influences whether reconnection spreads or convects with a fixed x-line length. (5) We perform a simulation initiating anti-parallel collisionless reconnection with a pressure pulse instead of a magnetic perturbation, finding spreading is unchanged rather than spreading at the magnetosonic speed as previously suggested. The results provide a theoretical framework for understanding spreading beyond systems studied here, and are important for applications including two-ribbon solar flares and reconnection in Earth's magnetosphere.

\end{abstract}
%\keywords{\textcolor{red}{See ApJ's list
%    (https://journals.aas.org/keywords-2013/) you are allowed up to
%    six} flares --- magnetic reconnection}
\maketitle

\section{Introduction}
\label{sec-intro}

Magnetic reconnection is a fundamental process that converts magnetic energy into kinetic and thermal plasma energy through a change in magnetic topology \citep{Dungey53,Vasyliunas75}. It mediates eruptive solar flares \citep{Priest00} and geomagnetic substorms \citep{McPherron73} and is thought to be an important process in numerous settings in high-energy astrophysics [{\it e.g.}, \citep{Uzdenski,Uzdensky16} and references therein]. Early models treated reconnection as two-dimensional (2D) \citep{Sweet58,Parker57,petschek64a}, but naturally-occurring reconnection is a 3D process [{\it e.g.}, \citep{Pontin11,Lukin11}]. 

One way the 3D nature of reconnection is manifested is that the x-line where the magnetic field topology changes has a finite extent in the direction normal to the plane of reconnection. [Note, we use the term "x-line" to refer simply to the line along which reconnection takes place, regardless of whether it is a separator, quasi-separator or a squashed 3D null point \citep{Priest14}. We do not use the term to mean a line of x-points, which can arise in 2D systems but is topologically unstable in 3D.] Spatially confined reconnection regions for which the extent of the region undergoing reconnection does not change in time have been studied theoretically and numerically (Refs.~\citep{Shay03,Linton06,Mayer13,Sasunov2015,Shepherd17,Liu19,Huang19} and Pyakurel et al., submitted). Alternately, the region undergoing reconnection can elongate, which we synonymously call spreading, over time. Such behavior has been observed in the solar corona during two-ribbon solar flares \citep{Isobe02,Lee08,Qiu09,Qiu10,Liu10ribbons,Cheng11,Tian15,Graham15,Qiu17} and prominence eruptions \citep{Tripathi06}, at Earth's magnetopause \citep{Zhou17,Zou18,Zou20}, in Earth's magnetotail \citep{McPherron73,Nagai82,Nagai13,Hietala14}, and in laboratory reconnection experiments \citep{Katz10,Egedal11,Dorfman13,Dorfman14}, and is thought to occur in the production of extremely extended reconnection events in the solar wind \citep{Phan06,Gosling07c,Shepherd17}.  Studying how reconnection spreads, which is the focus of the present study, is important in many settings because it impacts secondary processes such as particle acceleration and the global efficiency of the release of large-scale magnetic energy.

%\begin{comment}
%For the present study, we focus on two-ribbon solar flares, which show strong evidence of 3D spreading of magnetic reconnection. These types of solar flares are characterized by pairs of filamentary structures on the solar surface that appear to separate from each other in time and also spread (or elongate) along their length. The ribbon separation is understood via the CSHKP model of solar flares, in which the ribbons are produced by reconnection outflows that travel down the reconnected field lines to the loop footpoints, and successively reconnecting flux produces the apparent separation of ribbons \citep{Carmichael64,Sturrock66,Hirayama74,Kopp76}. The length-wise spreading, however, has undergone less scrutiny. The spreading of the ribbons at the solar surface is likely related to a spreading of the reconnection out of the plane of reconnection where it occurs in the corona causing the flare \citep{Isobe02}. Both unidirectional and bidirectional spreading has been observed in two-ribbon flares, and bidirectional spreading is correlated with strong out of plane magnetic (guide) fields \textcolor{red}{Why is this being stated?  If we keep it, we need to make sure the reader knows why we are telling them and we would need to include references}. Attempts to attribute the spreading to MHD waves have been made \citep{Kawaguchi82,Kitahara90}, however the spreading speeds of ribbons are much slower than the local Alfv\'en speed \citep{Qiu17}, so it is thought to spread as a result of the reconnection itself spreading.
%\end{comment}

There have been many numerical studies of 3D reconnection spreading in various settings. During anti-parallel quasi-2D reconnection, the consensus is that
%in a current sheet of an initial thickness $w_0$ larger than the ion inertial scale $d_i$, 
spreading occurs in the direction perpendicular to the reconnection plane at the speed and direction of the current carriers \citep{huba02,Huba03,Shay03,Karimabadi04,Lapenta06,Shepherd12,Nakamura12,Mayer13,Jain13}. If one species carries all the current, the spreading is unidirectional; if both species carry some current the spreading is bidirectional. %\uline{Geometrical effects in genuinely 3D systems can lead to different results, with a reconnection region that convects with the null point} \citep{Lukin11}. 

A number of physical mechanisms for spreading have been suggested.  It was argued \citep{huba02} that spreading of collisionless reconnection is caused by electrons convecting the reconnected magnetic into the region not undergoing reconnection [see also \citep{Hesse01,Shay03}]. They argued it was caused by a shock-like ``reconnection wave'' and motivated the result using linear theory. 
%propagates the reconnection, which extends the reconnecting region along the current sheet in the out of plane direction. 
The electron magnetohydrodynamic (eMHD) induction equation is
\begin{equation}
    \frac{\partial {\bf B}}{\partial t} = \frac{1}{ne} \nabla \times ({\bf J} \times {\bf B}),
\end{equation}
where ${\bf B}$ is the magnetic field, $n$ is the number density, $e$ is the proton charge, ${\bf J} = (c / 4\pi) \nabla \times {\bf B}$ is the current density, and $c$ is the speed of light.  Linearizing around the out of plane current profile, the perturbed reconnected (normal) magnetic field $B_{1n}$ is governed by
\begin{equation}
    \frac{\partial B_{1n}}{\partial t} + \frac{{\bf J}}{ne} \cdot \nabla B_{1n} = 0. \label{eq:conveqn}
\end{equation}
This shows that the magnetic field of the x-line is convected at a velocity associated with the current carriers, assumed to be electrons in their work.  When ions carry some of the current, spreading occurs at the speed of the current carriers in their respective directions \citep{Shay03,Lapenta06,Nakamura12}. If the thickness of the current sheet is $w_0$, then in the reference frame in which the electrons carry all the current, the spreading speed $v_s$ scales as 
%The empirical result can be quantified using a simple scaling analysis. From Amp\`ere's law (in Gaussian units), ${\bf J} = c\nabla\times{\bf B}/4\pi$, where ${\bf B}$ is the magnetic field and ${\bf J}$ is the current density.  Assuming the thickness of the reconnecting current layer is $\delta$, a scaling analysis gives $J \sim cB_0/4\pi \delta$, where $B_{0}$ is the reconnecting magnetic field strength. In the ion rest frame such that electrons carry the out-of-plane current, the electron velocity ${\bf v}_e = -{\bf J}/ne$. Then, the reconnection spreading speed $v_{s}$ in the out-of-plane direction is
\begin{equation}
  v_{s} \sim \frac{J}{ne} \sim \frac{cB_{up}}{4\pi n e w_0} = c_A \frac{d_i}{w_0}, \label{eq-vspreadl}
\end{equation}
where $B_{up}$ is the upstream reconnecting magnetic field strength, $d_i = c/\omega_{pi} = (m_ic^2 / 4\pi n e^2)^{1/2}$ is the ion inertial scale, $c_A = B_{up} / (4 \pi n m_i)^{1/2}$ is the Alfv\'en speed based on $B_{up}$, and $m_i$ is the ion mass.  The functional dependence on $w_0$ was confirmed in simulations \citep{Shay03}, and it was similarly shown that the relevant speed of the current carriers is that of the initial current sheet thickness $w_0$ rather than the kinetic-scale thickness after reconnection has started \citep{Lapenta06,tak}. Interestingly, if electrons carry all the current for anti-parallel reconnection, it was shown that reconnection does not spread in the resistive-MHD model  \citep{Nakamura12}. Moreover, reconnection can merely convect without the region undergoing reconnection elongating in the out of plane direction \citep{Shay03}.

An alternate mechanism for collisionless anti-parallel reconnection spreading was presented, based on pressure instead of magnetic field \citep{Huba03,Nakamura12}. The region where reconnection occurs was found to be of lower plasma pressure than the non-reconnecting regions.  The low pressure convects with the current carriers into the non-reconnecting regions, inducing inwards flow which causes reconnection sequentially in the out of plane direction. A related model was developed to explain observations of impulsive reconnection in the Magnetic Reconnection eXperiment (MRX) \citep{Dorfman13,Dorfman14}. In their experiment, the initial conditions had an electron flow gradient in the out of plane direction.  This gradient requires an inflow in an adjacent non-reconnecting region to preserve mass continuity, producing a sequential onset of reconnection. 

%Collisionless reconnecting current sheets with anti-parallel magnetic fields have a two-scale structure \citep{Sonnerup79}, with an ion diffusion region of thickness $d_i$ and an electron diffusion region of thickness $d_e$.  Distinguishing between them is not crucial for the present discussion, as it is sufficient to know these are small (kinetic) scales, so we refer to the thickness of the reconnection current %sheet as $w_0$. 

Spreading of magnetic reconnection is qualitatively different when there is a background out of plane (guide) magnetic field, which commonly is present in reconnection in solar flares \citep{Qiu17}, the solar wind \citep{Gosling05a}, and the dayside magnetopause \citep{Zou18}. Laboratory experiments showed that, for a strong guide field, the spreading is bidirectional with a speed given by the Alfv\'en speed $c_{Az} = B_{0z} / (4\pi m_i n)^{1/2}$ based on the guide field strength $B_{0z}$ rather than the speed of the current carriers \citep{Katz10}:
\begin{equation}
    v_s = \pm c_{Az}. \label{eq-vspreadguide}
\end{equation}
Two-fluid simulations found the same scaling with the out-of-plane (guide) magnetic field \citep{Shepherd12,Jain17}. The spreading has been described as being mediated by Alfv\'en waves \citep{Shepherd12}, whistler waves and flow induced waves \citep{Jain17}, and kinetic Alfv\'en waves \citep{tak}. 
%As the guide field strength is increased from zero, there is a transition from spreading with the speed and direction of current carriers to bidirectional spreading at the Alfv\'en speed based on the guide field \citep{Shepherd12}.  
Recently, it was shown in simulations of guide field reconnection with asymmetric plasma conditions that spreading in current sheets thinner than ion scales is bidirectional at the Alfv\'en speed, but is at the current carrier speed for thicker current sheets \citep{tak}. The different behavior for different current sheet thicknesses was attributed to the reduced tearing instability growth rate for wider current sheets. A guide field can also impede the spreading of reconnection due to the presence of multiple oblique x-lines \citep{Schreier10}.

This study presents a number of new results on the fundamental physics of the spreading of reconnection of finite extent. We generalize the theory of anti-parallel reconnection spreading \citep{huba02}, showing that it can be interpreted in the form of a scaling analysis and showing that the same theory can be used to derive from first principles the scaling of the spreading speed in the strong guide field limit, thereby uniting the understanding of reconnection spreading under a single first principles approach. 
%The theory is then used to develop a number of new insights about fundamental aspects of reconnection spreading: 
New results include: (1) We argue that the key physical aspect of x-line spreading is the induction of an x-line topology in the non-reconnecting region [see also Jain et al.~(2013)\citep{Jain13}], which is carried out by the gradient in the electric field in the outflow direction at the interface between the reconnecting and non-reconnecting regions. If an x-line topology is not induced in a given direction, reconnection does not spread in that direction. (2) The physical cause of reconnection spreading without and with a guide field are different, with convection at the Hall scale and MHD-scale magnetic field bending, respectively, playing key roles. We validate the theoretical results using 3D two-fluid and resistive-MHD numerical simulations, for both anti-parallel and guide field reconnection. (3) The theory explains why anti-parallel reconnection in the resistive-MHD model does not spread \citep{Nakamura12}. 
%(4) Conversely, it also provides an explanation of the spreading opposite to the electron convection direction \citep{Shay03}. 
(4) We find that a determining factor for whether a current sheet spreads or convects with a fixed length in a numerical simulation is the aspect ratio of the domain, which we suggest is controlled by the amount of free magnetic energy in the system. (5) Finally, we perform a test of whether the results obtained herein are dependent on the manner in which reconnection is excited in the system.  This is important because it has been suggested \citep{Vorpahl76} that reconnection spreads at the fast magnetosonic speed rather than the speed of the current carriers.  Physically, this mechanism could occur if reconnection is initiated through a pressure pulse squeezing the current sheet. As simulations typically initiate reconnection using a magnetic perturbation, it is important to assess whether the speed of the spreading depends on the way in which reconnection is seeded. Using a pressure pulse to initiate reconnection, we find that anti-parallel reconnection spreads at the current carrier speed rather than the magnetosonic speed.

The layout of this paper is as follows. In Sec.~\ref{sec-theory}, we discuss the theory of 3D reconnection spreading and derive a number of key implications about the physical cause of reconnection spreading and applications to collisionless and collisional systems with and without a guide field. In Sec.~\ref{sec-simulation}, we discuss our numerical simulation setup. In Sec.~\ref{sec-results}, we discuss the results of our simulations. 
%In Sec.~\ref{sec-app} we investigate what the results mean for two ribbon flares and other applications, and 
We offer conclusions and discuss new insights as a result of this work in Sec.~\ref{sec-conc}.

\section{Theory of reconnection spreading}
\label{sec-theory}

\subsection{General considerations}
\label{sec-theory-basic}

\begin{figure*}
%\figurenum{1}
\center  \includegraphics[width=6.8in]{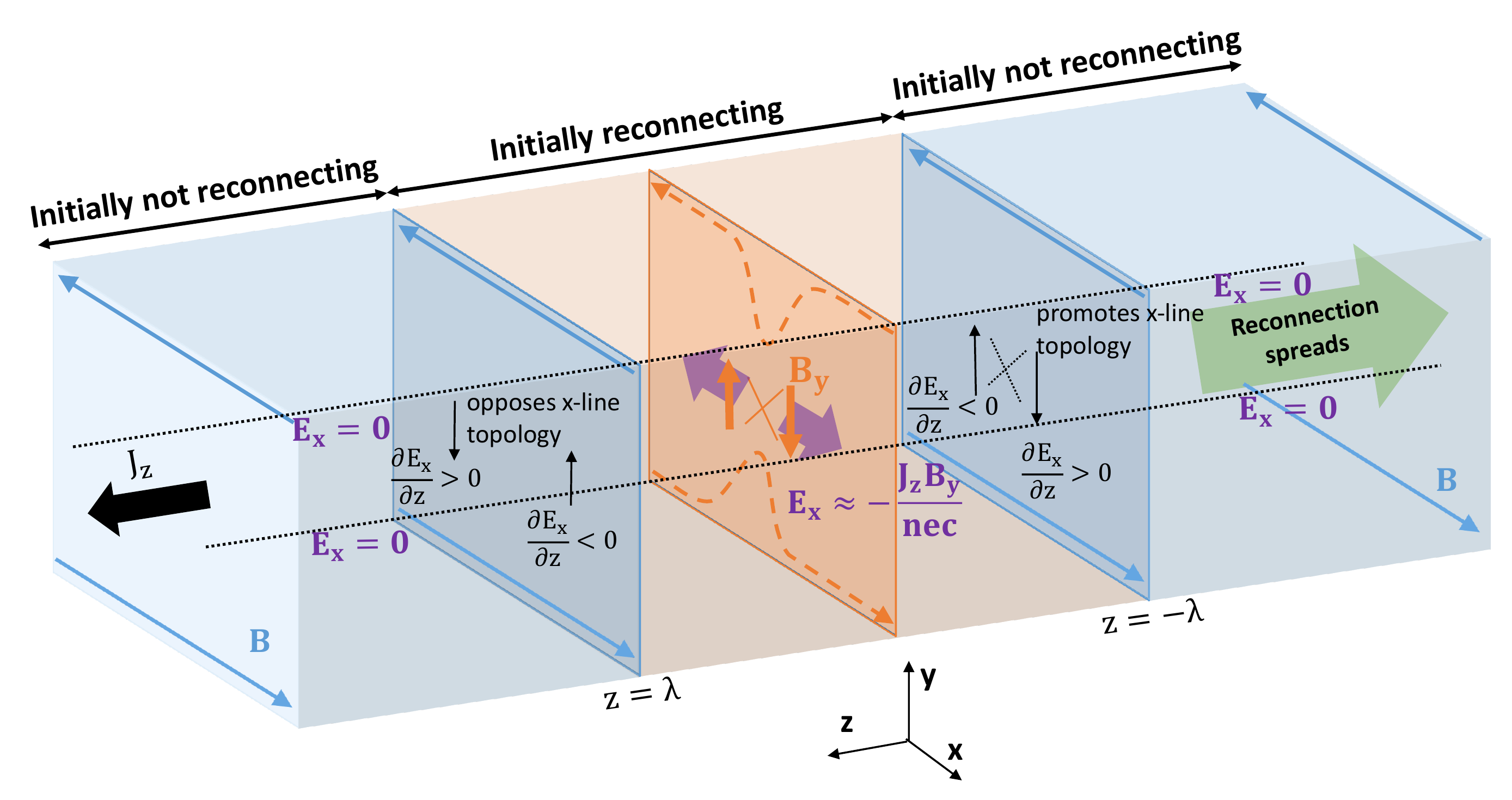}
%  \plotone{singlesheetrates.eps}
 % \plotone{curzmax2Dbig.eps}
  \caption{Sketch of a system undergoing anti-parallel reconnection in a localized region in the out of plane direction from $-\lambda < z < \lambda$, motivating the physics behind why reconnection spreads in the direction of electron convection. Orange shading denotes the finite domain where reconnection is taking place, and it is not taking place in the blue shaded region. The projection of a representative reconnecting magnetic field line in the orange region is shown in the $xy$ plane as the dashed orange line, with a dotted orange X denoting the x-line. The reconnected components of the magnetic field in the orange region are denoted by thick vertical orange arrows. %Associated with the reconnection in the orange shaded region, the Hall effect bends the orange magnetic fields in the $-z$ direction, associated with an $x$ component of 
  The Hall electric field component $E_x$ shown as purple arrows points away from the x-line in the reconnecting region and is zero elsewhere. The gradient in $E_x$ at the $z = -\lambda$ interface produces a normal magnetic field $B_y$ that promotes an x-line topology (black arrows), extending the x-line and causing spreading. At the $z = \lambda$ edge, the $B_y$ produced opposes an x-line topology (black arrows), so reconnection does not spread in that direction.}
  \label{fig-apmodel}
\end{figure*}

We use a coordinate system in which $z$ is the direction of the initial current, the current sheet is centered around $y = y_{cs}$, and $x$ is the direction of the equilibrium reconnecting magnetic field, with $B_x > 0$ for $y < y_{cs}$ and $B_x < 0$ for $y > y_{cs}$. We use a reference frame where the electrons fully carry the out of plane current. The asymptotic reconnecting magnetic field strength is $B_0$, there may be a guide field of strength $B_{0z}$, and the current sheet has an initial thickness $w_0$. The system is sketched schematically in Fig.~\ref{fig-apmodel} for the case without a guide field. The $y = y_{cs}$ plane is shown with two dotted lines. At a given time, reconnection is occurring in a localized part of the current sheet with finite out of plane extent $2\lambda$, shown with orange shading in the figure, while the parts of the system shaded blue are not undergoing reconnection.

As in the model reviewed in the Introduction, we consider the time evolution of the reconnected magnetic field $B_y$.  To generalize the previous approach \citep{huba02}, we begin from Faraday's law, $\partial {\bf B} / \partial t = - c \nabla \times {\bf E}$, where ${\bf E}$ is the electric field.
%, so understanding how $B_y$ is generated in the non-reconnecting portion of the current sheet requires understanding the spatial structure in ${\bf E}$.  
At the interface between reconnecting and non-reconnecting regions, gradients in the $z$ direction exceed gradients in the $x$ direction, so the term that dominates $B_y$ production is [see also \citep{Jain13}]
\begin{equation}
\frac{\partial B_y}{\partial t} \approx -c\frac{\partial E_x}{\partial z }. \label{induction}
\end{equation}
A scaling analysis allows us to find a characteristic out of plane spreading speed $v_s$, given by
\begin{equation}
v_{s} = \frac{\Delta z}{\Delta t} \approx -c\frac{\Delta E_x}{\Delta B_y },  \label{vs}
\end{equation}
where the spatial finite difference $\Delta z$ is evaluated at the boundary between the reconnecting and non-reconnecting regions, and we associate $\Delta z/\Delta t$ with the speed of the spreading $v_s$. We retain the minus sign, as it gives information about the direction of propagation. 

We first make contact with previous work.  For anti-parallel collisionless reconnection, it was argued that reconnection spreading occurs via out of plane convection by the electrons.  The electric field associated with this is $E_x \sim -J_z B_y / nec$, from the Hall term. Using this in Eq.~(\ref{vs}) and taking $\Delta B_y \simeq B_y$ at the interface between reconnecting and non-reconnecting regions gives
\begin{equation}
v_{s} \approx \frac{\Delta (J_zB_y/ne)}{\Delta B_y} \approx \frac{J_z}{ne}.
\label{uniform-speed}
\end{equation}
This reproduces the result that reconnection spreads at the speed and direction of the current carriers in Eq.~(\ref{eq:conveqn}). %\citep{huba02}.
In our approach, the result follows from a scaling analysis rather than linear theory.

In what follows, we argue that Eq.~(\ref{vs}) is useful for predicting the spreading speed beyond only anti-parallel reconnection.  More generally, the component of the net electric field in the outflow direction $E_x$ is given by the generalized Ohm's law 
\begin{equation}
E_x = -\frac{v_yB_z - v_zB_y}{c} + \frac{J_yB_z - J_zB_y}{nec} -\frac{1}{ne} \frac{\partial p_e}{\partial x} + \frac{m_e}{ne^2} \frac{d J_{x}}{dt} + \eta J_x, \label{ohmx}
\end{equation}
where ${\bf v}$ is the (ion) bulk flow velocity, $p_e$ is the (scalar) electron pressure, $m_e$ is the electron mass, and $\eta$ is the resistivity.  The right hand side includes the convection term, Hall term, electron pressure gradient term, electron inertia term, and resistive term in order of appearance. We show that in different settings, different terms can dominate.  We find the electron pressure gradient and electron inertia terms do not impact spreading in current sheets at or above ion inertial scale thicknesses. 
%We do not include collisions in our scaling analysis, but in Sec.~\ref{sec-additionaltests2} we address spreading in resistive MHD and argue that it is not a strong contributor, especially in most solar and magnetospheric environments where plasmas are only weakly collisional.  
%In the following subsections, we discuss spreading in anti-parallel and guide field reconnection.

Before considering specific systems, we elucidate what our approach reveals about the physical mechanism for reconnection spreading.  Previous work \citep{huba02} suggested the evolution of $B_y$ is what determines spreading. Physically, in order to seed an x-line in a plane in which there is initially no x-line, one needs to generate a normal magnetic field $B_y$ with a bipolar structure of the proper polarity [see also \citep{Jain13}]. If the $x$ coordinate of the x-line is $x^\prime$, then an x-line is seeded if $B_y > 0$ for $x < x^\prime$ and $B_y < 0$ for $x > x^\prime$ for the assumed $B_x$ directionality. From Eq.~(\ref{induction}), the signs of the gradient of the electric field $E_x$ at the ends of the region undergoing reconnection determine whether $\partial B_y / \partial t$ is locally positive or negative for $x < x^\prime$ and $x > x^\prime$, which determines whether an x-line develops over time in the non-reconnecting region. We argue the sign of the gradient of the electric field in the outflow direction is a more fundamental interpretation of how reconnection spreads via convection.

We note a subtlety that is important for numerical studies of reconnection spreading and may be important in naturally occurring reconnection. Many theoretical developments of reconnection spreading, including the treatment in this section, are based on the propagation of a small $B_y$ into regions not previously undergoing reconnection. However, the presence of $B_y$ is not synonymous with the onset of reconnection. Rather, the presence of $B_y$ triggers the tearing instability which makes $B_y$ grow in time, and it is only after getting to large amplitudes that steady reconnection is set up. Thus, there is a time delay between when $B_y$ spreads into a region not undergoing reconnection and when reconnection begins in earnest. This has been seen in previous simulation studies \citep{huba02}, and more recently has been noted as an important factor in the spreading of reconnection in thick current sheets for which the time scale of the tearing instability is longer \citep{tak}. For the present study, the time delay between the appearance of $B_y$ and the onset of full-fledged reconnection is the same at all locations, so the spreading speed of reconnection is unchanged by the delay. Consequently, in this study, it is sufficient to study the spreading speed of the normal magnetic field $B_y$ as a proxy for the spreading speed of the onset of full-fledged magnetic reconnection.
%The reconnecting field $B_y$ has previously been used as a diagnostic of reconnection spreading (e.g. \citep{Nakamura12,tak}).

%If the non-reconnecting current sheet develops an x-line topology, in the next instant in time it will similarly induce an x-line successively further out in $z$, extending the x-line in the out of plane direction and continuing the spreading. In time, the x-line in the non-reconnecting region goes unstable to tearing/reconnection, and reconnection is triggered.  Thus, there is a time delay between the seeding of an x-line and when reconnection begins in earnest.  Therefore, the x-line topology may spread on time scales shorter than the time scale required for the local current sheet to fully collapse to electron scales and reach a steady state reconnection rate. This has important implications for the spreading speed of anti-parallel reconnection.  In particular, it implies that the relevant speed of the current carriers for spreading is evaluated before reconnection starts, when the local current sheet thickness is $w_0$, not the speed of the current carriers once reconnection has started and the current sheet thins to electron scales. This is consistent with previous scaling results of anti-parallel reconnection, for which $v_s \propto 1/w_0$ \citep{Shay03,Shay04}.

\subsection{Spreading of collisionless anti-parallel reconnection}
\label{sec-uni}

We exploit the results of the previous subsection to develop new insight on the physics of spreading for anti-parallel collisionless (Hall) reconnection. In Fig.~\ref{fig-apmodel}, the dark blue arrows represent magnetic field lines within the blue shaded regions in which reconnection is not taking place, which are straight because there is no reconnection to bend them towards the $y = y_{cs}$ plane. In contrast, the dashed orange arrows depict the projection of a representative reconnecting magnetic field line in the $xy$ plane within the orange shaded region where reconnection is occurring, which bend in towards the x-line. As previously noted \citep{huba02,Shay03}, the out of plane current is carried in the $z$ direction by electrons convecting in the $-z$ direction. Thus, the x-line topology governed by $B_y$, depicted by the thick vertical orange arrows in the region undergoing reconnection, is convected in the $-z$ direction and reconnection spreads in that direction (the green arrow) in this reference frame.  
%The Hall field extends into the region in which reconnection is not occurring for $z = -\lambda$, seeding the x-line topology in the blue shaded region, so reconnection spreads in the $-z$ direction. This provides a physical understanding of why anti-parallel collisionless reconnection spreads in the direction of the current carriers in the non-reconnecting part of the sheet, in agreement with empirical results of past studies of anti-parallel reconnection spreading discussed in Sec.~\ref{sec-intro}. 

We reinterpret this in terms of the electric field $E_x$ and the induced magnetic field in the non-reconnecting regions. In the region where reconnection is taking place, the out of plane current $J_z$ (the thick black arrow) in the presence of the reconnected magnetic field $B_y$ (the thick orange arrows, negative for $x > x^\prime$, positive for $x < x^\prime$) produces a non-zero component of the Hall electric field in the outflow direction $E_x \approx -J_zB_y/nec$, pointing away from the x-line as denoted by the purple arrows.  This $E_x$ is relatively uniform between $z = -\lambda$ and $z = \lambda$, but at the boundaries of the reconnecting region at $z = \pm \lambda$, there is a non-zero out of plane gradient $\partial E_x / \partial z$.  From Faraday's law, this produces a $\partial B_y / \partial t$ in the adjacent non-reconnecting planes, positive for $x > x^\prime, z = \lambda$ and $x < x^\prime, z = -\lambda$ and negative for $x < x^\prime, z = \lambda$ and $x > x^\prime, z = -\lambda$, represented by the four thin black arrows in the $z = \pm \lambda $ planes. For the non-reconnecting plane adjacent to the $z = -\lambda$ boundary, there is initially no $B_y$, so the presence of a $\partial B_y / \partial t$ generates a magnetic field that seeds an x-line topology (shown as an X with black dotted lines), thus promoting the spreading of the x-line in the direction of electron convection, as expected. In contrast, in the non-reconnecting plane adjacent to $z = \lambda$, $\partial B_y / \partial t$ has the opposite polarity, which serves to weaken the existing $B_y$, and thus the x-line and therefore reconnection do not spread in the $+z$ direction.
%Thus, the x-line lengthens in the $-z$ direction into the non-reconnecting region.  Over time, this x-line topology goes unstable to reconnection, and the reconnection spreads progressively in the $-z$ direction, {\it i.e.,} the direction of electron convection.  
This provides an alternate, but equivalent, understanding of why reconnection does not spread in the direction of the current in the reference frame in which the electrons carry the current.

\subsection{Lack of spreading of anti-parallel reconnection in resistive-MHD}
\label{sec-eta}
In collisional reconnection described by resistive-MHD, the Hall, electron pressure, and electron inertia terms are dropped from the generalized Ohm's law [Eq.~(\ref{ohmx})].  Then, the only terms that can produce an $E_x$ are $-v_y B_z / c, v_z B_y / c$, and $\eta J_x$.  For anti-parallel collisional reconnection in the reference frame in which the electrons carry the out of plane current, $B_z$ and $v_z$ are both zero. Thus, there is no spreading in collisional reconnection due to convection in the reference frame in which the electrons carry the out of plane current. This result is consistent with previous resistive MHD simulations \citep{Nakamura12}, and provides a first-principles reason for the absence of spreading in this case.
%We briefly note a subtlety about the case in which ions carry some or all of the current. As has been pointed out previously, this is simply a change of reference frame from the case in which the electrons carry all the current \citep{Shay03,Lapenta06,Nakamura12}. If the ions carry some of the out of plane current, then the $v_z B_y / c$ term is non-zero.  However, $v_z$ is the same at both ends ($z = \pm \lambda)$ of the region undergoing reconnection, so {\it ion} convection causes both ends of the reconnecting region to convect. However, they convect at the same speed $v_z$, so the reconnection region convects without spreading in this reference frame, consistent with the result of simply changing the reference frame.
We point out that the resistive term, with $E_x = \eta J_x$, can in principle cause spreading. This spreading is bidirectional, as magnetic diffusion of $B_y$ at the boundary between the reconnecting and non-reconnecting region induces an x-line topology in the non-reconnection part.  Using $\Delta E_x = \Delta (\eta J_x) \sim \eta c B_y / 4 \pi L_{z0}$, where $L_{z0}$ is the length scale of the transition between the reconnecting and non-reconnecting regions, Eq.~(\ref{vs}) gives
\begin{equation}
    |v_s| \sim \frac{\eta c^2}{4 \pi L_{z0}}, \label{eq-diffusionspread}
\end{equation}
{\it i.e.,} the diffusion velocity across the boundary. This mechanism may be relevant for spreading in collisional plasmas, such as the chromosphere or some laboratory experiments.  However, for most settings of heliophysical interest, the resistivity is exceedingly small, so the spreading due to resistivity is small on dynamical time scales.

\subsection{Spreading of guide field reconnection}
\label{sec-component}

\begin{figure*}
%\figurenum{2}
\center  \includegraphics[width=6.8in]{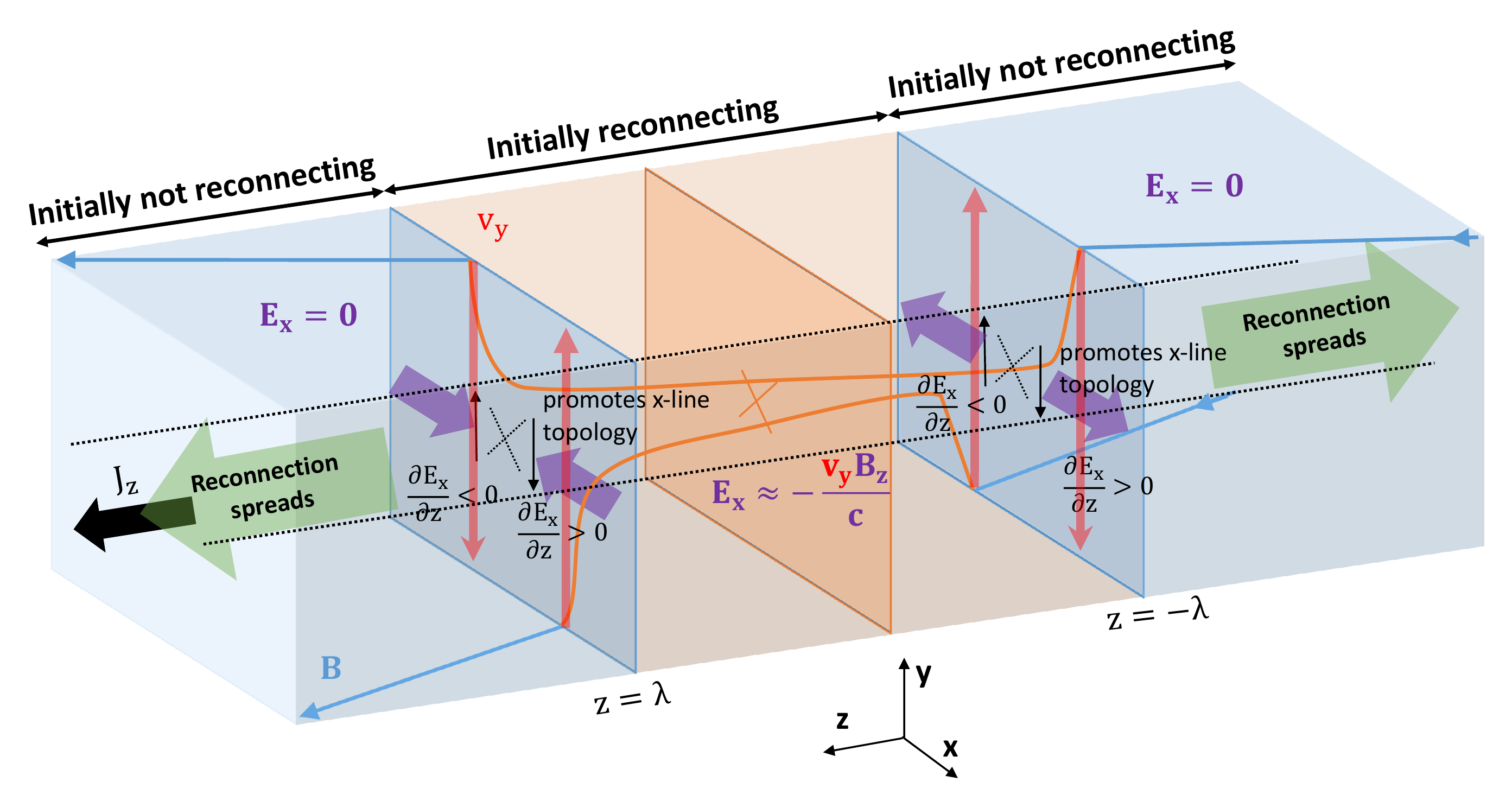}
%  \plotone{singlesheetrates.eps}
 % \plotone{curzmax2Dbig.eps}
  \caption{Analogously to Fig.~\ref{fig-apmodel}, sketch of a system undergoing guide field reconnection in a localized region in the out of plane direction from $-\lambda < z < \lambda$, motivating the physics behind why guide field reconnection spreads bidirectionally.  Reconnection between $-\lambda < z < \lambda$ convects the magnetic field towards the neutral line, which sets up a strongly bent magnetic field at the interfaces.  This strong magnetic curvature drives flow in the normal direction (red arrows).  This flow immersed in the guide field sets up an $E_x$, shown as the purple arrows.  The gradient in $E_x$ at the interfaces produces x-lines at both the $\pm \lambda$ boundaries, extending the x-line in both directions.}
  \label{fig-gfmodel}
\end{figure*}

The physical cause of reconnection spreading with a non-zero guide field $B_{0z}$ is fundamentally different than with no guide field.  For simplicity, we consider the limit where the guide field is much larger than the reconnecting magnetic field, $B_{0z} \gg B_0$, and that the current sheet is not sufficiently thicker than the ion gyroscale, at which point the mechanism for reconnection spreading can change because of the guide field dependence of guide field reconnection on the growth rate of the collisionless tearing instability \citep{tak}. 
%Numerical studies on the scaling of component reconnection with guide field strength have shown that for $B_{0z}\geq 3B_0$, component reconnection transitions into a bidirectional spreading mode at the Alfv\'en speed based on the guide field, that is $v_s = \pm c_A_z = \pm B_{0z} / (4 \pi n m_i)^{1/2}$ \citep{Shepherd12}. 
Figure~\ref{fig-gfmodel} shows a sketch of reconnection spreading in the large guide field limit. As in Fig.~\ref{fig-apmodel}, blue regions are not initially undergoing reconnection.  The magnetic field in this region is shown with dark blue lines, depicted with a strong $z$ component.  The region initially undergoing reconnection with length $2\lambda$ is shown in orange. As reconnection occurs, the upstream magnetic field in this region convects inward towards the neutral line, shown in dark orange lines. This bends the upstream magnetic field, introducing a kink in the magnetic field localized near the interface of the reconnecting and non-reconnecting regions.  This kinked magnetic field provides a curvature force, which drives a bulk flow in the vertical ($\pm y$) direction, depicted as red arrows near the $z = \pm \lambda$ planes. The flow, therefore, has a quadrupolar structure in the $xz$ plane.  The normal flow in a region with a guide field produces a convective electric field $E_x \approx -v_yB_z/c$, depicted by the purple arrows. It is strongest in a thin region near the interface, and also has a quadrupolar structure in the $xz$ plane.  Since $E_x$ has a gradient in the $z$ direction, Faraday's law implies that $B_y$ is generated in that region, with the sign of $B_y$ being given by Eq.~(\ref{induction}).  The induced $B_y$ fields are depicted by the thin black arrows at $z = \pm \lambda$.  At both edges, the magnetic topology generated by the induced $B_y$ is of an x-line, so the x-line spreads in both out-of-plane directions. This is consistent with the known result that guide field reconnection spreads bidirectionally.  We stress that this sketch of the physics is valid for both collisionless and collisional reconnection.  Essentially, this bending of the magnetic field line is physically similar to a rotational discontinuity or launching an Alfv\'en wave. For collisionless reconnection in current sheets at gyroscales, this field line bending becomes a kinetic Alfv\'en wave, as was previously elucidated \citep{tak}.

We now perform a scaling analysis to obtain the spreading speed in the strong guide field limit. The bulk flow in the $y$ direction due to the curvature force from the bent upstream magnetic field is described by the momentum equation
\begin{equation}
\frac{\partial v_y}{\partial t} \approx \frac{B_{0z}}{4\pi m_i n}\frac{\partial B_y}{\partial z }. \label{curvature}
\end{equation}
In writing this, we use that the large guide field limit implies $v_yB_{0z} \gg v_zB_y$ in the convection electric field in Ohm's law, as both the out of plane bulk flow $v_z$ and the reconnected field $B_y$ are small during the early stages of reconnection. 

A scaling analysis on this equation gives 
\begin{equation}
v_y \sim \frac{B_{0z}\Delta B_y}{4 \pi m_i n v_s},  \label{eq-vyscale}
\end{equation} 
where we have taken $\Delta v_y \sim v_y$ between adjacent reconnecting and non-reconnecting planes and $v_s=\Delta z/\Delta t$ as per equation~(\ref{vs}).
%In the large guide field limit, for early times when x-line spreading takes place before the current sheet collapses down to electron inertial scales, the convective electric field dominates over the Hall electric field in equation~(\ref{ohmx}). 
Then, the relevant term in Eq.~(\ref{ohmx}) gives $E_x$ as 
\begin{equation}
E_x \approx -\frac{v_yB_{0z}}{c} 
\sim -\frac{B_{0z}^2\Delta B_y}{4 \pi m_i n v_s c}, \label{ohmx-guide}
\end{equation}
%This is not possible in a purely anti-parallel field configuration, where the dominant term in equation~(\ref{ohmx}) is the Hall field which has a bipolar structure for $E_x$, as discussed in Sec.~\ref{sec-uni}. 
and using this result in Eq.~(\ref{vs}) reveals
\begin{equation}
v_{s} \approx \pm \left(\frac{B_{0z} ^2}{4\pi m_i n}\right)^{1/2} = \pm c_{Az}.
\label{guide-speed}
\end{equation}
This reproduces the known result that reconnection with a large guide field spreads bidirectionally at the Alfv\'en speed based on the guide field $c_{Az}$ in Eq.~(\ref{eq-vspreadguide}).

\section{Simulation Setup}
\label{sec-simulation}

%\subsection{3D Setup}
Simulations are carried out using the two-fluid code F3D \citep{Shay04}, which updates the continuity, momentum, induction, and pressure equations, and can include the Hall, resistive, and electron inertia terms in the generalized Ohm's law.  Time is stepped forward using the trapezoidal leapfrog algorithm \citep{Guzdar93} and spatial derivatives are fourth order finite differences.  For simulations with the Hall term, lengths are normalized to the ion inertial scale $d_{i0} = (m_ic^2/4\pi n_0 e^2)^{1/2}$, time is normalized to the inverse ion cyclotron frequency $\Omega_{ci0}^{-1}=m_ic/eB_0$, velocities to the Alfv\'en speed $c_{A0}=B_0/\sqrt{4\pi m_in_0}$, electric fields to $c_{A0} B_0 / c$, and temperatures to $m_i c_{A0}^2 / k_B$, where $B_0$ is the initial upstream reversing magnetic field magnitude, $n_0$ is the initial upstream density, and $k_B$ is Boltzmann's constant. When the Hall term is absent, the only differences to the normalizations are that lengths are normalized to an arbitrary length $L_{MHD}$, times are normalized to $L_{MHD}/c_{A0}$, and resistivity is normalized to $4 \pi c_{A0} L_{MHD} / c^2$.

The computational domain has dimensions $L_x \times L_y \times L_z = 102.4 \times 51.2 \times 256.0$, where $x$ and $y$ correspond to the outflow and inflow directions in 2D, respectively, and $z$ is perpendicular to the 2D reconnecting plane. Boundary conditions are triply periodic, and the system size is chosen to be large enough that the boundaries do not impact the relevant dynamics. The grid scale is $\Delta x \times \Delta y \times \Delta z = 0.05 \times 0.05 \times 1.0$. The lower resolution in the out-of-plane direction has been used before \citep{Shay03,Shepherd12}, and is justified since out-of-plane dynamics in our setup change more slowly than dynamics in the reconnection plane. When electron inertia is included, the ion-to-electron mass ratio is $m_i/m_e=25$, and we expect the relevant results are independent of this value, since previous work on the spreading of anti-parallel reconnection has shown that the dynamics of x-line spreading are insensitive to the mass ratio \citep{Shay03} and the terms in Ohm’s law that contribute to reconnection spreading in the theory (the Hall and convection electric fields) are independent of the mass ratio.

The initial conditions consist of two oppositely directed current sheets. The $x$-component of the initial magnetic field is given by $B_{0x}=\tanh[(y + L_y/4)/w_0] - \tanh[(y-L_y/4)/w_0)] - 1$, where $w_0$ is the initial current sheet thickness. When a guide field $B_{0z}$ is included, it is uniform.  The initial density is uniform with a value of 1, and a non-uniform temperature varying from 1 to 1.5 is used to balance magnetic pressure in the current sheet. The plasma pressure is provided fully by ions and is treated as adiabatic, while the electrons are assumed cold at all times. The electrons carry all of the initial current.

The resistivity $\eta$ is identically zero for simulations employing the Hall term, and is 0.004 for the resistive-MHD simulations. Fourth-order diffusion is included in all equations with coefficients $D_{4x} = D_{4y} = 1.6\times 10^{-5}$ and a larger diffusion coefficient in the $z$ direction $D_{4z}=1.6\times 10^{-1}$ because the grid scale is larger. The time step is $0.02$ for all simulations with no guide field. For simulations with a guide field, a smaller time step of $0.01$ and a larger fourth-order diffusion coefficient $D_{4z}=6.4\times 10^{-1}$ are used to account for the faster dynamics in the out of plane direction. The diffusion coefficient and time step values are varied in trial simulations to ensure they do not play any significant role in the numerics.

%\begin{figure} 
%\figurenum{3}
%%\center  %\includegraphics[width=3.4in]{initialconditions.eps}
%\center \includegraphics[width=3.4in]{2Dcurzplot.eps}
%\center \includegraphics[width=3.4in]{2Dpertplot.eps}
%  \caption{Representative initial conditions for a simulation. The plots give a cut in the $yz$ plane at $x = -L_x/2$ of (a) the equilibrium current $J_z$ with $w_0 = 1$, and (b) the $y$-component of the magnetic perturbation $B_{1y}$ of amplitude 0.005 on only the upper current sheet. \label{fig-initconds}}
%\end{figure}

We employ simulations with initial thickness $w_0 = 1.0$. We repeat the simulation of anti-parallel reconnection with different uniform current sheet thicknesses $w_0 = 0.5, 2.0$ and $3.0$ and the guide field simulation with $w_0 = 2$ to confirm that in all cases, the local physics remain qualitatively similar to their $w_0 = 1$ counterparts and that the spreading speeds are consistent with previous work \citep{Shay03}.
%For anti-parallel reconnection in non-uniform current sheets, we employ additional simulations with an initial current sheet thickness profile that varies in the out of plane direction between two specified values $w_1$ and $w_2$, given by
%\begin{equation}
%w_{0}(z) = \frac{1}{2} \left\{\left( w_1 + w_2 \right) + \left( w_1 - w_2 \right) \left[\tanh \left(\frac{z + L_{0}}{w_z}\right) - \tanh \left(\frac{z - L_{0}}{w_z}\right) - 1\right]\right\}, \label{eq-wofz}
%\end{equation}
%where $w_1$ and $w_2$ are the current sheet thicknesses at equilibrium in the perturbed and unperturbed regions, respectively, $L_0 = 80\ d_{i0}$ is the half-length of the perturbed region, and $w_z = 4\ d_{i0}$ is the gradient scale. We carry out four simulations holding $w_1 = 1.0di_{i0}$ fixed and varying $w_2 = 1.25, 1.5, 1.75, 2\ d_{i0}$ and an additional two simulations holding $w_2 = 2.0d_{i0}$ fixed and varying $w_1 = 0.75, 1.25 d_{i0}$. 
%Thicker initial current sheets are desirable but, because they take longer to evolve, are significantly more computationally expensive.
When a guide field is included, we use $B_{0z} = 3.0$, which is sufficient to be in the large guide field limit. 
%{\bf Figure~\ref{fig-initconds}(a) shows initial conditions for the out-of-plane current $J_z$ in a 3D simulation with $w_0 = 1$ in a cut in the $yz$ plane at $x=-L_x/2$.}

%/
%     ep0 = ep_norm*alx_tot/(4*3.141592654)

%          fzz(k) = 0.5 + 0.5*(tanh((zz(k) + l0pert)/w0z) -tanh((zz(k) - l0pert)/w0z) -1.)
 %          if (jj(j) <= ny0d2p3 ) then
 %             epv(i,j,k,3)= 0!ep0*(1+cos(pi* (yy(j)+aly_tot/4.)*4./aly_tot)) &
 %                       ! * sin(pi/2* xx(i)*4/alx_tot)*fzz(k)
 %          else
 %             epv(i,j,k,3)= ep0*(1+cos(pi*(yy(j)-aly_tot/4.)*4./aly_tot)) &
%                        * sin(pi/2* xx(i)*4/alx_tot)*fzz(k)
%     call curl (epv, bf_pre)

Unless otherwise stated, we initialize the simulations with a coherent perturbation in the magnetic field. To do so, the $z$ component of the perturbed vector potential is defined as
\begin{equation}
A_{1z}(x,y,z) = \frac{\tilde{B}_{1}L_x}{4 \pi}\left[1 + \cos\left(\frac{4\pi(y-L_y/4)}{L_y} \right) \right] \sin\left(\frac{2\pi x}{L_x}\right) f(z) \label{eq-pert}
%A_{1z}(x,y,z) = \left\{
%        \begin{array}{ll}
%\frac{\tilde{B}_{1}}{4 \pi L_y}\left[1 + \cos\left(\frac{4\pi(y-L_y/4)}{L_y} \right) \right] \sin\left(\frac{2\pi x}{L_x}\right) f(z) & y\geq 0 \\ 0 & y < 0
%        \end{array}
%    \right., \label{eq-pert}
\end{equation}
for $y \geq 0$ and 0 for $y < 0$, where $\tilde{B}_{1} = 0.005$ is a constant and the envelope $f(z)$ has the form
\begin{equation}
f(z) = \frac{1}{2} \left[\tanh \left(\frac{z + w_{0pert}}{2}\right) - \tanh \left(\frac{z - w_{0pert}}{2}\right)\right], \label{eq-fofz}
\end{equation}
where $w_{0pert}$ defines the initial half-extent of the coherent perturbation in the out of plane direction. We use $w_{0pert} = 15$ unless stated otherwise. The resulting magnetic perturbation ${\bf B}_1 = -{\bf \hat{z}} \times \nabla A_{1z}$ seeds an x-line/o-line pair in the $xy$ plane for only the upper current sheet at $y = y_{cs} = L_y/4$, localized to $-w_{0pert} < z < w_{0pert}$. 
%Figure~\ref{fig-initconds}(b) shows a representative plot of $B_{1y}$. 
%For simulations with non-uniform current sheet thickness, we choose $w_{0pert}$ in equation~(\ref{eq-pert}) to ensure the perturbation is localized exclusively in the region of lower thickness $w_1$ so that any reconnection observed in the region of larger thickness $w_2$ is due to spreading of reconnection and not due to the initial perturbation. 
We perturb only the upper current sheet because doing so prolongs the timescale for the interaction between the two current sheets due to flows in the $y$-direction and thus ensures the reconnection occurring in the upper sheet at later times is purely due to reconnection spreading in the upper sheet. 
%The key features we are studying are unchanged in simulations with both sheets perturbed (not shown).  
To ensure the spreading has no dependence on $w_{0pert}$, we perform a suite of simulations with an anti-parallel field configuration with $w_0 = 1$ with varying $w_{0pert}$ of $9, 12, 15,$ and $30$, with all other parameters held the same. We find that $w_{0pert}$ affects the initial extent of the reconnection region in the $z$ direction at the time of onset, which is to be expected, but the spreading of reconnection is unaffected. Incoherent noise in the $x$ and $y$ components of the magnetic field at the $10^{-5}$ level is also used to break symmetry, which prevents secondary magnetic islands from staying in the initial x-line location \citep{Shay04}. 

\begin{figure*}
%\figurenum{3}
\center  \includegraphics[width=6.8in]{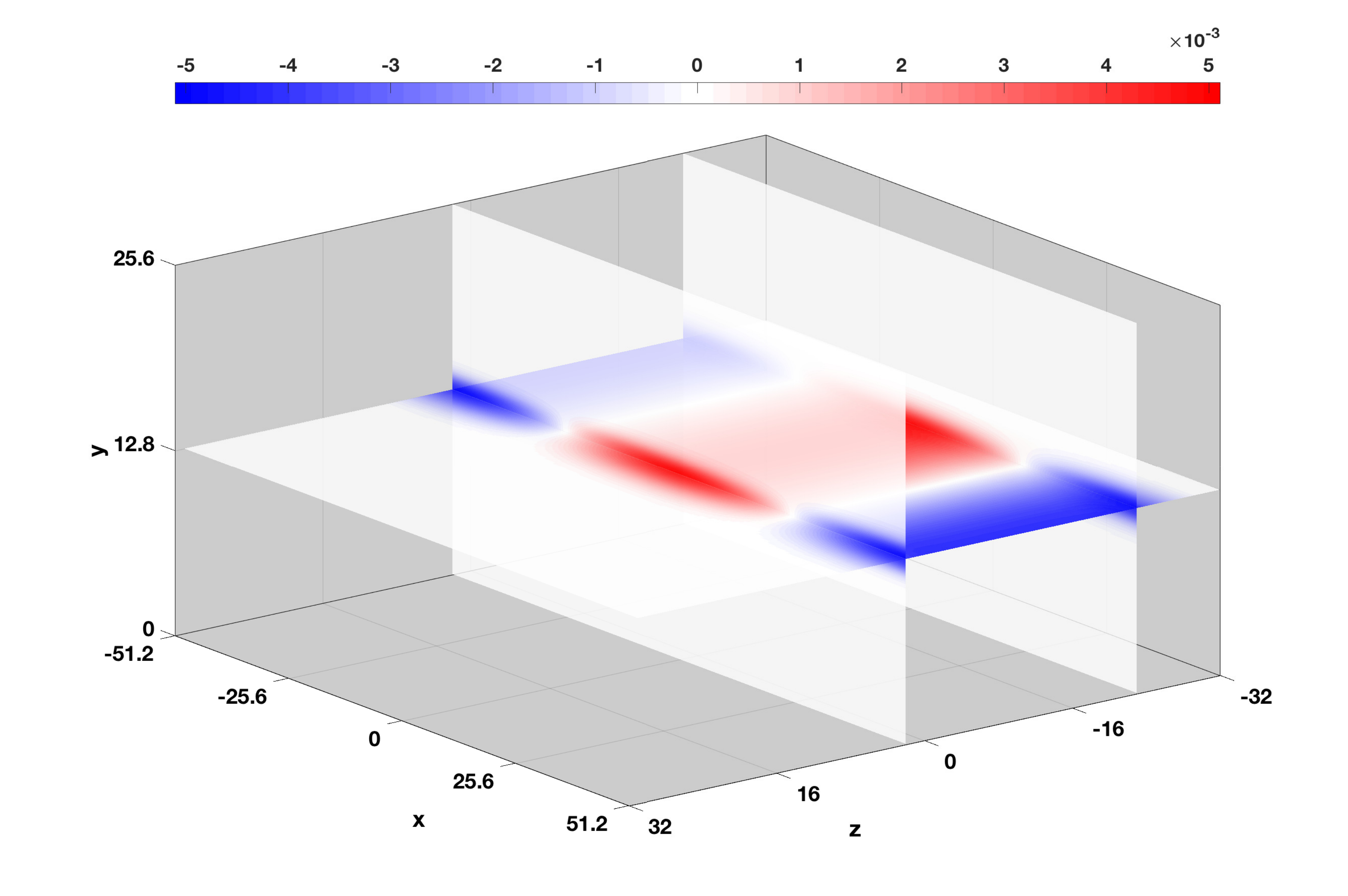}
  \caption{Electric field $E_x$ from a 3D simulation of anti-parallel reconnection with a current sheet of initial thickness 1 at $t = 10$. Data are plotted at planes of $y = y_{cs} = 12.8$, $z = 2$, and $z = -20$.  The x-line is at $x = -25.6$, and the o-line is at $x = 25.6$. The structure of $E_x$ is consistent with the sketch in Fig.~\ref{fig-apmodel}, motivating why anti-parallel reconnection spreads only in the direction of the current carriers.} \label{fig-m3}
\end{figure*}

%We repeat these simulations with the magnetic perturbation on both current sheets instead of only the top one and also for a different system size with $L_x \times L_y = 51.2 \times 51.2$. 
%In both cases, the scaling of the onset time with the initial sheet thickness remains qualitatively similar.

\section{Results}
\label{sec-results}
\subsection{Anti-parallel collisionless reconnection spreading}
\label{sec-results-u}
%\begin{comment}
%We begin with an overview of the simulation results for uniform thickness current sheets to validate our method of determining the reconnection spreading speed before applying it to non-uniform current sheets. \end{comment}

We begin by testing the theory in Sec.~\ref{sec-uni}. To do so, we need to verify the structure and the dominant contributor of the electric field $E_x$ and the time evolution of the magnetic field $B_y$ near the boundary of the initial reconnecting region.  
%For anti-parallel reconnection, we use the simulation with a uniform thickness $w_0 = 1$. 

%In this simulation, fast reconnection is not observed anywhere until $t = 70$.
%, similar to the onset timescale in the 2D benchmark simulation with equal current sheet thickness discussed in Sec.~\ref{sec-benchmark}. %We examine an early time $t = 10$ before reconnection onset to test the assumptions of the theory.

%We investigate the electric and magnetic field structure at $t = 10$, when the x-line is spreading but before the reconnection is going at its steady rate.
We investigate the electric and magnetic field structure at $t = 10$, when the perturbed $B_y$ is spreading out from its initial location, but before full-fledged reconnection is going at its steady rate (which occurs closer to $t = 80$), as discussed in Section~\ref{sec-theory-basic}. Figure~\ref{fig-m3} shows the net electric field component $E_x$. Planar cuts through the upper (perturbed) current sheet $y = y_{cs} = 12.8$ and through the reconnecting planes at $z = 2$ and $z = -20$ near the boundaries of where reconnection occurs are shown. These $z$ planes are selected because the x-line seeded by the perturbation is initially between $z = \pm 15$, but this region drifts in the $-z$ direction, the direction of electron convection. The initial convection of the perturbed region before reconnection spreads is consistent with the behavior observed in Fig. 1 of Huba and Rudakov (2002)\citep{huba02}, though it was not discussed in their study. 
%, so the initial x-line and ultimately fast reconnection begins not in $-w_{0pert} < z < w_{0pert}$, but further in the negative $z$ direction. 

The red-white-blue color map for the electric field $E_x$ ranges from $-0.005$ to $0.005$. The two zeroes of $E_x$ in the reconnecting region (in white) at $x = x^\prime = -25.6$ and $x = 25.6$ coincide with the x-line and o-line, respectively. The electric field $E_x$ is qualitatively similar at later times when the x-line is significantly longer in extent in the $-z$ direction.
%and is no longer merely convecting.
The bipolar structure of $E_x$ points outwards from the x-line, consistent with the sketch in Fig.~\ref{fig-apmodel}. The largest contributor to $E_x$ is the Hall term $E_x^{Hall} = -J_zB_y/nec$, as expected \citep{huba02}. It has a maximum magnitude of 0.005 and is two orders of magnitude larger than the next largest contribution from the convection term $v_zB_y/c$. 
%This is consistent with the model in Section~\ref{sec-uni}. 
 
\begin{table*}
\center
    \begin{tabular}{|c|c|c|c|c|}
        \hline
        ~Anti-Parallel                                 & $(x=-38.4,2<z<10)$ & $(x=-38.4,-28<z<-20)$ & $(x=-12,2<z<10)$ & $(x=-12,-28<z<-20)$ \\ \hline
        $-\frac{\partial E^{Hall}_x}{\partial
        %commented stuff is for net Ex:
        %z}$ & -0.00045            & 0.00045              & 0.00048         & -0.00048             \\ 
        z}$ & -0.00045            & 0.00045              & 0.00048         & -0.00048             \\ 
        $\frac{\partial B_y}{\partial t}$ & -0.00056           & 0.00058               & 0.00056          & -0.00056            \\
        \hline
    \end{tabular}
       \caption{Comparison of the out of plane gradient of the Hall electric field $-\partial E^{Hall}_x / \partial z$ and the local time derivative $\partial B_y / \partial t$ for the $w_0 = 1$ anti-parallel collisionless reconnection simulation at $t=10$. The spatial gradient is measured over the specified ranges in $z$ near the boundaries of the reconnecting region, and the time derivative is measured between $t=9$ and $11$ at the midpoint of the specified ranges in $z$. The agreement confirms that the electrons convect the x-line topology in the $-z$ direction for anti-parallel reconnection. \label{table-ap}}
\end{table*}

\begin{table*}
\center
    \begin{tabular}{|c|c|c|c|c|}
        \hline
        ~ Guide Field 3                                 & $(x=-38.4,38<z<48)$ & $(x=-38.4,-48<z<-38)$ & $(x=-12,38<z<48)$ & $(x=-12,-48<z<-38)$ \\ \hline
        $-\frac{\partial E^{conv}_x}{\partial %z}$ & 0.00037             & 0.00034               & -0.00040          & -0.00036            \\ 
        z}$ & 0.00050             & 0.00026               & -0.00053          & -0.00027            \\ 
        $\frac{\partial B_y}{\partial t}$  & 0.00041             & 0.00035               & -0.00043           & -0.00038            \\
        \hline
    \end{tabular}
    \caption{Same as Table~\ref{table-ap}, but for the convective electric field at the given locations and times in the guide field $3$ case. The agreement confirms that the convection electric field gradients propagate the x-line topology in the $\pm z$ directions for guide field reconnection.} \label{table-gf}
\end{table*}

To quantitatively confirm that the induction of $B_y$ is caused by $E_x^{Hall}$, we compute the out of plane gradient $\partial E^{Hall}_x/\partial z$ at both boundaries of the reconnecting region to the left and right of the initial x-line, $x=-38.4$ and $x=-12$, respectively. We use least squares to fit a line to $E_x^{Hall}$ as a function of $z$ through the center line of the upper current sheet from $-28<z<-20$ and $2<z<10$ to approximate $\partial E_x^{Hall} / \partial z$. We calculate the local time derivative $\partial B_y/\partial t$ at the midpoint of the specified ranges in $z$, determined with a time-centered difference between $t=9$ and $11$. The results are gathered in Table~\ref{table-ap}. The similarity between the two terms shows that the main contribution to $\partial B_y / \partial t$ comes from $-\partial E_x^{Hall} / \partial z$, as expected. 

The signs of $\partial B_y / \partial t$ in the $z = -24$ plane are negative at $x=-12$ and positive at $x=-38.4$, to the left and right of the zero of $B_y$, respectively. This serves to promote an x-line topology in the $z = -24$ plane. In contrast, at $z = 6$, the signs of $\partial B_y/\partial t$ oppose the formation of an x-line, consistent with our explanation of why reconnection does not spread in the $z$ direction. These results confirm our theoretical predictions for anti-parallel collisionless reconnection spreading. 

\subsection{Guide field collisionless reconnection spreading}

\begin{figure*}
%\figurenum{4}
\center  \includegraphics[width=6.8in]{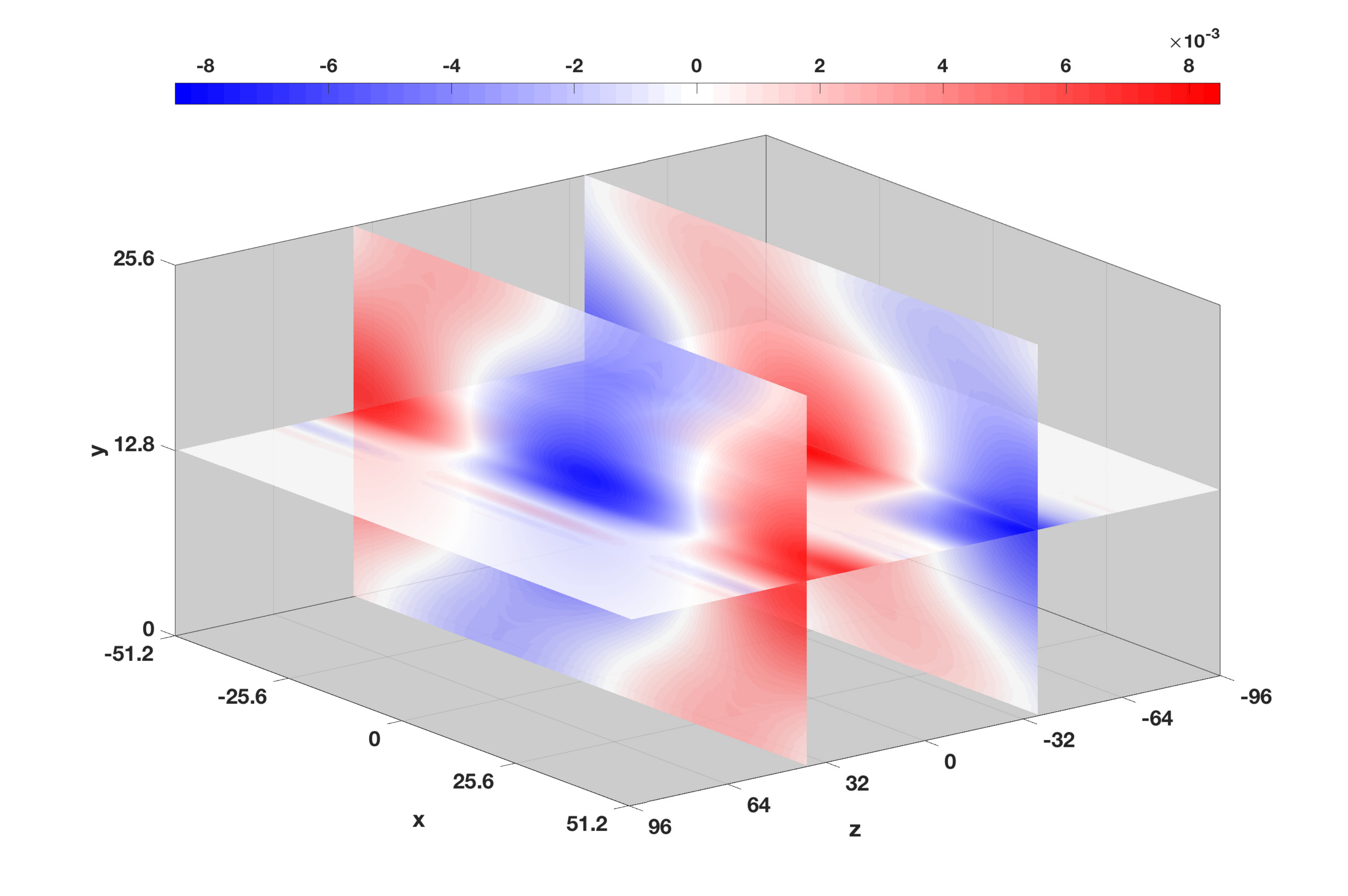}
  \caption{Electric field $E_x$ from a 3D simulation of guide field reconnection with a guide field $3$ with a current sheet of initial thickness $w_{0} = 1$ at $t = 10$. Data are plotted at planes of $y = y_{cs} = 12.8$, $z = -38$, and $z = 38$.  The x-line is at $x = -25.6$, and the o-line is at $x = 25.6$. The structure of $E_x$ is consistent with the sketch in Fig.~\ref{fig-gfmodel}, motivating why reconnection spreads bidirectionally.} \label{fig-mg3}
\end{figure*}

We now discuss the large guide field case. Figure~\ref{fig-mg3} shows the net electric field component $E_x$ at $t = 10$ for a simulation with guide field $B_{0z} = 3$, again when the perturbed $B_y$ is spreading out from its initial location, but before full-fledged reconnection is going at its steady rate. Planar cuts are shown at $y = y_{cs}$ through the upper (perturbed) current sheet and the $z = -38$ and $z = 38$ planes near the boundaries between the reconnecting and non-reconnecting regions. The two zeroes of $E_x$ at the intersections of the planes (in white) are again located at approximately $x = x^\prime = -25.6$ and $x = 25.6$, coinciding with the x-line and o-line, respectively. Note, $E_x$ has a much larger extent in the $y$ direction than for the anti-parallel case, which is localized within an ion inertial scale. That $E_x$ extends far beyond the current sheet to MHD scales is typical of reconnection with a large guide field. When reconnection is occurring, ion inflows $v_y$ extend into the upstream region, well outside the current layer to MHD scales. In the absence of a guide field $B_{0z}$, the associated convection electric field $E_x \sim v_yB_{z}/c$ in the upstream region is negligible. This is because the only contribution to $B_z$ is the quadrupolar Hall magnetic field, which is very small at MHD scales upstream of the diffusion region. This explains why $E_x$ is localized to the current layer in the case without a guide field (see Fig.~\ref{fig-m3}). However, if there is a large guide field $B_{0z}$, the electric field $E_x \sim v_yB_z/c$ is non-zero, both at Hall scales and beyond the current layer into MHD scales because of the ion inflow, as is seen in Fig.~\ref{fig-mg3} %This is consistent with the guide field introducing MHD-scale physics, as predicted. 

The largest contributor to the electric field $E_x$ during the initial spreading phase is the convection term $E_x^{conv} = v_yB_z/c$, which has a maximum magnitude of 0.009 and is three times larger than than the next largest contribution from $E_x^{Hall} = J_zB_y/nec$. The quadrupolar structure of $E_x$ points inwards towards the x-line at $z = 38$ and outwards from the x-line at $z = -38$, consistent with the sketch in Fig.~\ref{fig-gfmodel}. 

We note that there is a small amplitude oscillatory signature at the leading edges of the $E_x$ signal. This is reminiscent of low amplitude oscillatory behavior observed in Jain and Buchner’s (2017) study\citep{Jain17} in the outermost edges of the reconnecting region, although in our simulation we do not see the larger amplitude oscillations they observed in between. Understanding these differences is outside of the scope of the present study. Regardless, due to the smallness of this oscillatory signal in our study ($\sim0.001$ compared to $\sim 0.008$ for the non-oscillatory signal), it is not playing any significant role in the spreading. To quantitatively confirm the induction of $B_y$ is caused by the convective electric field, we compute $\partial E^{conv}_x/\partial z$ at both boundaries of the reconnecting region to the left and right of the initial x-line, $x=-38.4$ and $x=-12$ respectively, using a similar approach as the previous subsection, at $-48<z<-38$ and $38<z<48$. We then compare to $\partial B_y/\partial t$ at the midpoint of the specified ranges $z=-43$ and $z=43$, computed as in the previous subsection. The results are gathered in Table~\ref{table-gf}. The similarity between the two quantities shows that the main contribution to $\partial B_y / \partial t$ comes from $-\partial E_x^{conv} / \partial z$, as we predict in our scaling of Eq.~(\ref{ohmx-guide}) for the large guide field limit. The signs of $\partial B_y / \partial t$ show that $B_y$ develops with a negative sign at $x=-38.4$ and a positive sign at $x=-12$, to the left and right of the zero of $B_y$, respectively, at both $z = -42$ and $z = 42$. This implies the magnetic topology is that of an x-line at both $z=-42$ and $z=42$ planes, consistent with the model for why the x-line spreads in both the $z$ and $-z$ directions.

Additionally, we test the prediction that the bulk flow $v_y$ is driven by the curvature force due to the bent upstream magnetic field at the boundaries of the reconnecting region by directly computing the left and right hand sides of Eq.~(\ref{curvature}) with simulation data. For the right hand side, we first compute $\partial B_y / \partial z$ at $t=10$ near the ends of the reconnecting region through $y = y_{cs}$, using a least squares fit of $B_y$ as a function of $z$ from both $-42<z<-35$ and $35<z<42$ and to the left and right of the x-line at $x=-38.4$ and $x=-12.8$, respectively. Then, the numerical estimate for the curvature force term is $(B_{z}/n)\partial B_y / \partial z$ (in code units). The left hand side $\partial v_y / \partial t$ is then computed locally at the midpoints $z = 38$ and $z = -38$ with a time-centered difference between $t=9$ and $11$. The results are gathered in Table~\ref{table-cf}. The similarity of the two terms is strong evidence that the curvature force drives the bulk flows $v_y$ near the boundaries of the reconnection region in guide field reconnection. In summary, our simulation results confirm the predictions about the electric and magnetic fields in guide field reconnection in  Section~\ref{sec-component}.

\begin{table*}
\center
    \begin{tabular}{|c|c|c|c|c|}
        \hline
        ~ Momentum Eqn.                                   & $(x=-38.4,35<z<42)$ & $(x=-38.4,-42<z<-35)$ & $(x=-12,35<z<42)$ & $(x=-12,-42<z<-35)$ \\ \hline
        $\left(\frac{B_z}{n}\right)\frac{\partial B_y}{\partial z}$ & -0.00021            & 0.00024               & 0.00026           & -0.00043            \\ 
        $\frac{\partial v_y}{\partial t}$    & -0.00021            & 0.00014               & 0.00023           & -0.00045            \\
        \hline
    \end{tabular}
    \caption{Comparison of the curvature force term $(B_z/n)\partial B_y/\partial z$ (in code units) and the local acceleration of the bulk flow $\partial v_y / \partial t$, in the guide field reconnection spreading simulation at $t=10$.
    The spatial derivative is averaged over the given ranges in $z$ near the boundaries of the reconnecting region, and the time derivative is determined from a time-centered difference between $t=9$ and $11$ at the midpoint in the specified ranges in $z$. This confirms that the curvature force drives the vertical flows.} \label{table-cf}
\end{table*}

\subsection{Spreading of collisional reconnection in resistive-MHD}
\label{sec-additionaltests2}

Here, we study the spreading of collisional reconnection in resistive-MHD to test the predictions in Sec.~\ref{sec-eta}. For these simulations, the initial out-of-plane length scale in $z$ of the transition is $L_{z0} = 2$ from Eq.~(\ref{eq-fofz}) and $\eta= 0.004$, so Eq.~(\ref{eq-diffusionspread}) gives a predicted spreading speed of $v_s \simeq 0.002$ (in code units). For the duration of the simulations carried out here, the distance reconnection would spread from resistive effects is expected to be negligible. Consequently, we expect no spreading for anti-parallel reconnection, but spreading will occur for guide field reconnection. 

We carry out two 3D simulations using the resistive-MHD model. One has no guide field, and one has guide field $B_{0z} = 3$. The initial current sheet thickness is $w_0 = 0.32$. All other system properties and initialization parameters are the same as described in Sec.~\ref{sec-simulation}. For the anti-parallel reconnection simulation, we find that reconnection does not spread up to the simulated time $t = 200 \ L_{MHD}/c_{A0}$ (not shown). If spreading were to occur at the speed of the current carriers $v_s = J_z/ne  = 1/w_0 \approx 3 \ c_{A0}$, we would expect reconnection would spread a distance $\approx 600 \ L_{MHD}$ in the simulated time. This would be clearly observable, as this is longer than the computational domain in the $z$ direction. The region undergoing reconnection also does not convect in the out of plane direction, as the electrons initially carry all the current. These findings are consistent with previous results \citep{Nakamura12}. The region undergoing reconnection also does not convect in the out of plane direction, as the electrons initially carry all the current.  With $B_{0z} = 3$, reconnection spreads bidirectionally (not shown), as in the collisionless two-fluid simulation.  This confirms that, for the parameters of our study, there is no spreading for anti-parallel reconnection in which electrons carry the current within the resistive-MHD model, that MHD physics drives the spreading when there is a guide field, and that collisions play no important role in reconnection spreading.

\subsection{Dependence of spreading on system aspect ratio}
\label{sec-results-nu}

%We now test the spreading speed prediction for anti-parallel reconnection in current sheets with non-uniform thicknesses in equation~(\ref{nonuniform-speed}). First, we discuss how we find where reconnection is taking place and how we determine the speed at which the reconnection is spreading in our 3D simulations, since the flux function approach to find the reconnection rate employed in Sec.~\ref{sec-benchmark} does not rigorously work in 3D. 
%\begin{comment}
%\figurenum{4}
%\center   \includegraphics[width=7.2in]{allplots.eps}
%  %\plotone{bydiffspeed335.eps}
%  %\plotone{bydiffspeed334.eps}
%  %\plotone{bydiffspeed336.eps}
%  %\plotone{bydiffspeed330.eps}  
%  \caption{Average reconnected magnetic field $\tilde{B_y}(z,t)$, defined in equation~\ref{eq-byavgdef}, plotted as a function of the out-of-plane direction $z$ and time $t$, for initial current sheet thicknesses $w_0$ of (a) 0.5, (b) 1, (c) 2, and (d) 3 $ \ d_{i0}$. Black triangles denote when $\tilde{B_y} = 0.04$ is met for a chosen range in $z$, which we use as the condition for reconnection onset.  The out-of-plane reconnection spreading speed $v_s$, listed for each simulation, is the slope of these points determined from a least squares fit. \label{fig-stack}}
%\end{comment}

An interesting result arises from comparing our anti-parallel collisionless reconnection simulation with $w_0 = 2$ (used to confirm the results from the $w_0 = 1$ simulation do not depend on the current sheet thickness) with previous knowledge. In particular, we find that reconnection spreads with $w_0 = 2$, while a previous study found that reconnection with a current sheet of that thickness developed a reconnecting region of finite extent in the $z$ direction and simply convected at the speed of the current carriers rather than spread \citep{Shay03}; see the dashed lines of their Fig.~3a. Similar behavior was observed in other studies of Hall reconnection spreading in relatively thick current sheets \citep{Mayer13}.

%In Fig.~\ref{fig-scalingu}, we gather the spreading speeds for the anti-parallel reconnection simulations with uniform current sheet thicknesses calculated with the same methodology described in Sec.~\ref{sec-results-nu}, shown as a function of the initial current sheet thickness $w_0$. The prediction of the spreading speed from equation~(\ref{uniform-speed2}) is shown as a dashed line, showing excellent agreement between simulations and theory. It also agrees with a prior numerical study by \citet{Shay03}. 

The difference between the present and previous simulations is that the prior studies used a square computational domain in the $xy-$plane, whereas our domain is twice as big in the $x$ direction than in the $y$ direction. 
%They found the current sheet with a thickness of $2 \ d_{i0}$ developed an x-line with a fixed length that convected at the speed of the local current carriers \citep{Shay03}. 
%This differs from our study, where $2 \ d_{i0}$ and $3 \ d_{i0}$-thick current sheets develop an x-line that continually spreads at the same speed. 
We repeat our simulation with a current sheet of thickness $w_0 = 2$ in a square computational domain with $L_x \times L_y \times L_z = 51.2 \times 51.2 \times 256.0$, as in Shay et al.~(2003)\citep{Shay03}. We also find the x-line remains a fixed length and convects rather than spreads. 

We demonstrate our result graphically using a plot of the reconnected magnetic field $B_y$.
%The strength of the normal magnetic field component $B_y$ near the reconnection region is an indicator of the presence of reconnection. 
For a given $xy$ plane we find the reconnection region by first finding the zeroes of $B_y$ through the symmetry line of the current sheet in the $y = y_{cs} = L_y/4$ plane and determine if the magnetic topology is that of an x-line or an o-line.  For a current in the $z$ direction, $B_y$ changing from positive to negative with increasing $x$ is an x-line and from negative to positive is an o-line. If there are multiple x-lines, we define the primary one as that with the largest out-of-plane current $J_z$. The strength of the reconnected field $B_y$ increases from zero away from the reconnection region until the downstream edge of the electron diffusion region.  At every plane of constant $z$ and at every time $t$, we use the average magnitude of $B_y$ at the left and right downstream edges of the electron diffusion region as a proxy for the appearance of reconnection and denote this quantity as $\tilde{B}_y(z,t)$.

For the distance from the x-line to the downstream edges of the electron diffusion region, we note that the collisionless reconnection rate $E$ is typically $\sim 0.1$ [{\it e.g.}, \citep{shay99a,Birn01,Liu17}], which is also comparable to the aspect ratio of the diffusion region $\delta / L$, where $\delta$ is its thickness (in the $y$ direction) and $L$ is its length (in the $x$ direction). The thickness of the electron diffusion region \citep{Vasyliunas75} is the electron inertial scale $d_e$, so one expects the length $L$ of the electron diffusion region to be approximately $10 \ d_e$. For this study, $L \approx 2 \ d_{i0}$ since $d_e = 0.2 \ d_{i}$.  We validate this choice by visual inspection of cuts of $B_y$, confirming this choice of $L$ reasonably represents where the strength of the reconnected magnetic field is near its first maximum. Consequently, we compute the average reconnected field magnitude $\tilde{B}_y(z,t)$ at the downstream edges of the electron diffusion region for each $xy$ plane as
\begin{equation}
 \tilde{B}_y(z,t) = \frac{\left|B_y\left(x^\prime +
   L,y_{cs},z,t\right)\right| + \left|B_y\left(x^\prime -
   L,y_{cs},z,t\right)\right|}{2}, \label{eq-byavgdef}
 \end{equation}
where $x^\prime$ is the location of the x-line in the plane in question. 
%We note the x-line/o-line pair does not necessarily remain in on the symmetry line ($y=L_y/4$ here) if there is a guide field [{\it e.g.}, \citep{Grasso07,Daughton11b,Shepherd12,Baalrud12}] or a reconnecting magnetic field or density asymmetry \citep{Hoshino83,Scholer89,LaBelleHamer95,Siscoe02,Dorelli04,Cassak07d}, so this technique would need to be modified in such cases.

%We validate this approach in our 2D simulations where the average reconnected magnetic field magnitude $\tilde{B}_y$ is easily compared to the reconnection rate $E$ obtained through the flux function approach. We expect $\tilde{B}_y$ in equation (\ref{eq-byavgdef}) to increase in time from 0 to an asymptotic value of $\simeq$ 0.1 when it reaches a quasi-steady state. We directly compare $\tilde{B}_y(t)$ and $E(t)$ for all of the 2D anti-parallel simulations discussed in Sec.~\ref{sec-benchmark}.  Excellent agreement is found for the onset times (when both quantities exceed $0.04$). This gives us confidence that we can use this diagnostic in 3D simulations as a proxy for the local reconnection rate. 

\begin{figure}
%\figurenum{5}
\center  
  \includegraphics{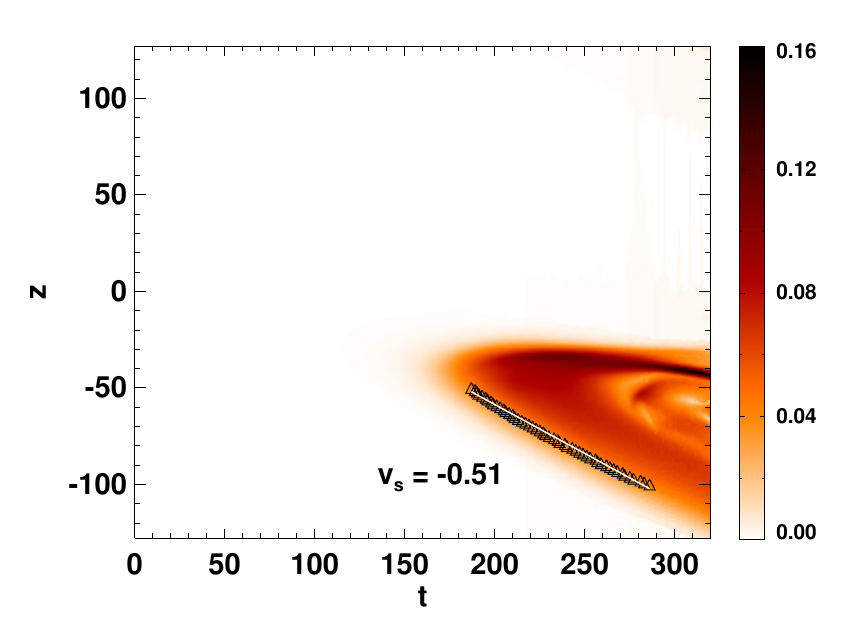}
  \caption{Average reconnected magnetic field $\tilde{B_y}(z,t)$, defined in Eq.~(\ref{eq-byavgdef}), as a function of out of plane position $z$ and time $t$, for the anti-parallel reconnection simulation with $w_0 = 2$. The triangles mark where $\tilde{B}_y$ crosses over 0.04, and the white line gives the best fit of these points, giving the spreading speed $v_s$.} This plot shows reconnection spreads, in contrast to the results in a square computational domain in which the x-line convects without spreading [Fig.~3a of Shay et al.~(2003)\citep{Shay03}]. \label{fig-stack336}
\end{figure}

The average reconnected field $\tilde{B}_y(z,t)$ for the upper current sheet $y = y_{cs} = L_y/4$ is shown as a stack plot of time $t$ and out-of-plane coordinate $z$ for the rectangular computational domain in Fig.~\ref{fig-stack336}. The time delay between the spreading of $B_y$ and the onset of full-fledged reconnection is apparent as the white space on the left side of the plot; the reconnection does not begin at $z = -45$ until $t \approx 170$ even though the perturbed $B_y$ would have reached that location by $t = 60$.
%each have $w_1 = 1\ d_{i0}$ in the thinner part of the current sheet, but are varied with $w_2 = 1.25, 1.5, 1.75,$ and $2.0 \ d_{i0}$ in the thicker part.
%Panels (a) through(d) are for different initial current sheet thicknesses $w_0$ of $0.5, 1, 2$ and $3 \ d_{i0}$, respectively.  
%Each horizontal cut represents data from a fixed $xy$ plane as a function of time $t$, while each vertical cut represents the spatial extent of $\tilde{B}_y$ in the $z$ direction of the reconnecting region at a fixed time. 
The triangular shape of $\tilde{B}_y(z,t)$ is representative of reconnection that is spreading, and not convecting with a fixed extent. A reconnecting x-line that is merely convecting without spreading would appear as a diagonal stripe, with a fixed extent in the $z$-direction. An example of this is in Fig. 3(a) of Shay et al.~(2013)\citep{Shay03}; the dashed lines there show a reconnecting x-line that convects with a fixed length. The interior structures within the overall triangular region are due to magnetic islands that arise after reconnection at that location has reached its steady state. To calculate the spreading speed, we define onset at a given $xy$ plane to be when $\tilde{B}_y$ exceeds 0.04. Onset times for individual $xy$ planes are plotted in Fig.~\ref{fig-stack336} as black triangles for a chosen interval in $-100<z<-50$. The spreading speeds are simply the slope of the collection of points denoting the onset time. We determine this slope using a least squares fit. We find the spreading speed is $v_s \approx -0.51$, consistent with Eq.~(\ref{eq-vspreadl}), as expected. Moreover, in a separate simulation with $w_0 = 3$ with the same square domain, we find that fast reconnection does not occur in the simulated time, whereas reconnection does occur in the rectangular cross section domain and spreads at the current carrier speed rather than merely convecting. 

%run226(w0=3) and run227
These results suggest that the aspect ratio of the reconnecting plane of the computational domain contributes to whether reconnection spreads or convects.  We hypothesize that reconnection in the square domain stops because the reconnection runs out of free magnetic energy relatively quickly, while in the rectangular domain there is more free energy and thus the reconnection persists longer, consistent with the stack plot. This is only a single simulation, though, so future work is needed to test this hypothesis.

\subsection{Dependence on perturbation structure}
\label{sec-additionaltests3}

To test whether the results obtained here are dependent on the manner in which reconnection is initiated in the system, we carry out an anti-parallel reconnection simulation with current sheet thickness $w_0 = 1$, with the Hall effect and electron inertia turned on.  Instead of perturbing the current sheet with a coherent perturbation of the magnetic field [see Eq.~(\ref{eq-pert})], we perturb the system with localized regions of higher plasma pressure localized just upstream of the current sheet. This perturbation is designed to drive flow towards the current sheet in a localized region to seed an x-line. The pressure perturbation $P_1$ we employ has the form 
\begin{eqnarray}
P_1(x,y,z) = P_{i1} \exp\left\{-0.5\left[ \left(\frac{x-x^\prime}{w_{0x}}\right)^2 + \right. \right. \\ \left. \left.  \left(\frac{y-y_{cs} - y_{0pert}}{w_{0y}}\right)^2 + \left(\frac{y-y_{cs} + y_{0pert}}{w_{0y}}\right)^2\right]\right\}f(z), \nonumber
\end{eqnarray}
where $P_{i1} = 2$ is the amplitude of the upstream pressure perturbation, $x^\prime = -L_x/4$ is the desired $x$ coordinate of the x-line, $w_{0x} = 4$ is the extent of the pressure perturbation in the $x$ direction, $y_{cs} = L_y/4$ is the $y$ location of the center of the current sheet being perturbed, $w_{0y} = 1$ is the thickness of the pressure perturbation in the $y$ direction, $y_{0pert} = 4$ is the distance upstream of the current layer on each side that the pressure perturbation is centered, and $f(z)$ is the same envelope in Eq.~(\ref{eq-fofz}) enforcing localization in $z$ between $\pm w_{0pert}$.

\begin{figure}
\center  
\includegraphics[width=3.4in]{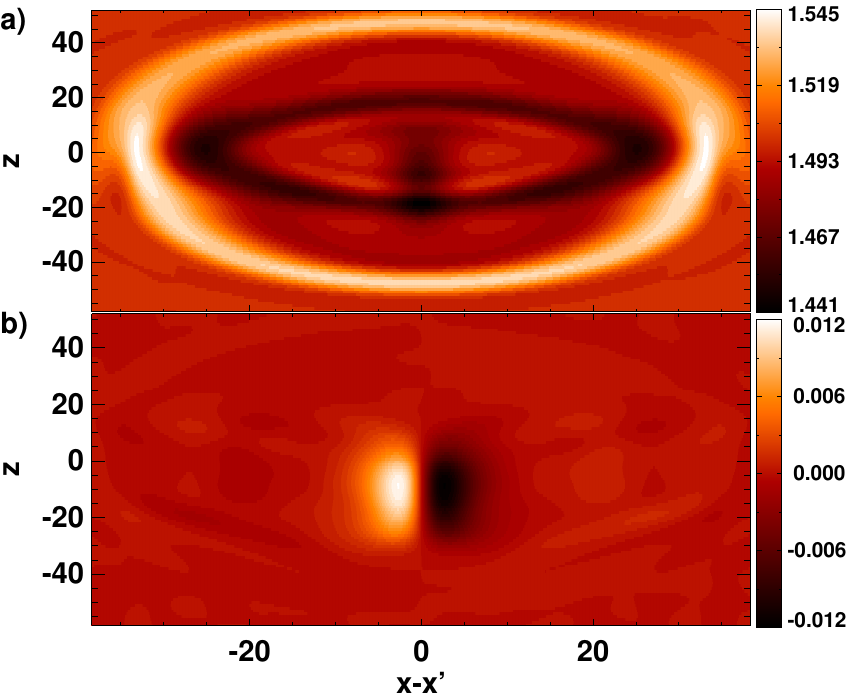}
  \caption{Planar cuts through $y = y_{cs}$ from a 3D simulation of anti-parallel Hall reconnection with a current sheet of initial thickness $w_0 = 1$ at $t = 20$. This simulation seeds reconnection with a pressure pulse instead of a magnetic perturbation. (a)  Ion pressure $P_i$ and (b) reconnected magnetic field $B_y$. The pressure pulse propagates bidirectionally at the fast magnetosonic speed, but reconnection spreads in the direction of the current carriers in the current sheet. \label{fig-run410pfi}}
\end{figure}

\begin{figure}
%\figurenum{1}
\center  
\includegraphics[width=3.4in]{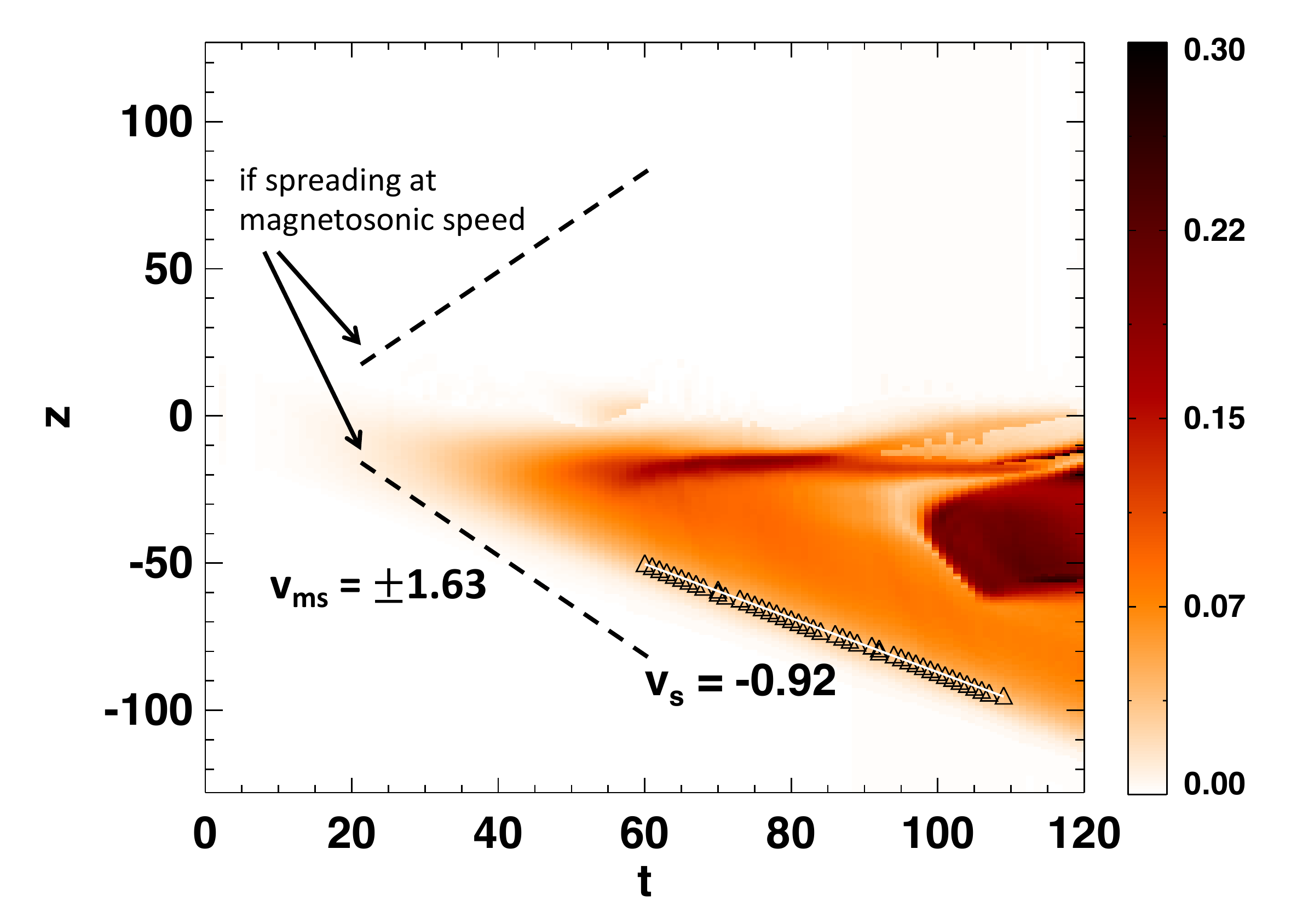}
  \caption{Average reconnected magnetic field $\tilde{B_y}(z,t)$, defined in Eq.~(\ref{eq-byavgdef}), as a function of out of plane position $z$ and time $t$, for the anti-parallel reconnection simulation with $w_0 = 1$ initiated with a pressure perturbation. Dashed lines show the expected x-line bounds if spreading occurs bidirectionally at the fast magnetosonic speed $v_{ms}$. Instead, spreading occurs unidirectionally with electron flow speed $v_s$ in the current sheet, denoted by the white line. %This plot shows that anti-parallel Hall reconnection spreads unidirectionally in the direction and with the approximate speed of the electrons in the current sheet $v_s$, not bidirectionally at the fast magnetosonic speed $v_{ms}$, represented with dashed lines, associated with the expanding pressure perturbation, consistent with the theory.
  }
  \label{fig-run410stack}
\end{figure}

%{\bf We find that the the onset of fast reconnection happens earlier than $t=70 \ \Omega_{ci}^{-1}$, the time a current sheet of the same thickness perturbed with the coherent perturbation in the magnetic field $B_y$ onsets.} 
The pressure perturbation launches a pulse that propagates out in all directions. The ion pressure $P_i$ in a cut through $y = y_{cs}$ at $t = 20$ is shown in Fig.~\ref{fig-run410pfi}(a). The leading edge of the pressure pulse at this time is at $z \approx \pm 50$, so its velocity is approximately 35 / 20 $\approx 1.75$.  This compares favorably to the fast magnetosonic speed, $v_{ms} = (c_A^2 + c_s^2)^{1/2} = (1 + 5/3)^{1/2} = 1.63$, as expected. The pressure pulse drives flow toward the current sheet which initiates reconnection, as desired, and the reconnection does spread in time.  The reconnected magnetic field $B_y$ in the same plane at the same time is shown in Fig.~\ref{fig-run410pfi}(b). The figure shows that reconnection spreads unidirectionally in the direction of the current carriers, not bidirectionally.  The average reconnected field $\tilde{B}_y(z,t)$ at $y = y_{cs}$ is shown as a stack plot in Fig.~\ref{fig-run410stack}. Using the same method described in Sec.~\ref{sec-results-nu}, the reconnection spreading velocity is found to be $\approx -0.92$, represented as a white line.  This is consistent with the velocity of the current carriers, which is -1 for this simulation. For reference, two dashed black lines are sketched representing what the boundaries of the structure in the stack plot would be if reconnection were to spread bidirectionally at the fast magnetosonic speed $v_{ms} \approx \pm 1.63$, showing clear disagreement.  

We conclude that perturbing the current sheet using a coherent perturbation in the magnetic field does not introduce a bias in our theoretical model for anti-parallel reconnection spreading, at least for the simulation presented here. More research is needed to see whether there are scenarios in which reconnection can spread with the fast magnetosonic speed, as has been previously suggested \citep{Vorpahl76}.

\section{Discussion and Conclusions}
\label{sec-conc}

We study the out-of-plane spreading of 3D magnetic reconnection that begins localized in the out of plane $z$ direction. We build off of a previously developed analysis \citep{huba02} for the spreading of anti-parallel collisionless reconnection. It describes the unidirectional spreading as caused by electrons convecting the reconnected magnetic field out of the reconnection plane (which they dubbed a ``reconnection wave''), and quantified by the Hall term in the electron-MHD induction equation.

In this study, we re-envision the previous model using a scaling analysis of Faraday's law using the full generalized Ohm's law rather than a linear analysis. We show that the same analytical approach can be used in a unified manner to describe the spreading of collisionless reconnection with a guide field, resulting in the known spreading speed given by the Alfv\'en speed and spreading bidirectionally \citep{Katz10,Shepherd12}.  The same approach also provides an explanation for why anti-parallel reconnection in resistive-MHD does not spread \citep{Nakamura12}.

Importantly, this new interpretation provides an alternate, first-principles understanding of the mechanism of reconnection spreading. In this approach, reconnection spreading occurs when the x-line extends, which requires the seeding of an x-line in the non-reconnecting region [see also Jain et al. (2013)\citep{Jain13}]. Thus, if the normal ($y$) component of the magnetic field is induced in a way that produces an x-line topology, then reconnection spreads. The normal magnetic field subsequently grows in time due to the tearing instability [see also Li et al. (2020)\citep{tak}] until steady-state reconnection is reached, with some time delay after $B_y$ enters the region previously not undergoing reconnection. If no x-line topology is induced at the interface between reconnecting and non-reconnecting regions, then the x-line does not spread in that direction. The induction of the normal magnetic field is controlled by the out of plane gradient of the component of the electric field in the outflow direction ($x$).  For anti-parallel reconnection in which electrons carry the current, the Hall term dominates the $E_x$ contribution due to electron convection.  For reconnection with a strong guide field, spreading is caused by the bending of the upstream magnetic field, which occurs predominantly at the boundary between the reconnecting and non-reconnecting regions. This sets up a strong magnetic curvature force driving flows in the $y$ direction. There is an associated electric field $E_x$ due to convection $v_y$ through the guide field $B_z$, which induces the necessary $B_y$ to produce an x-line topology in both directions. This mechanism relies only on MHD physics, showing that collisional guide field reconnection can spread similar to collisionless guide field reconnection.

The scaling analysis has other important implications. It reveals that collisions can cause reconnection to spread bidirectionally at the diffusion speed, which could be important for collisional reconnection in the laboratory and some settings in the Sun. However, in most settings where collisionality is quite weak, collisions play no important role in reconnection spreading.
%as the the time scale for diffusion in the out of plane direction is much larger than the time scales for reaching the steady state reconnection rate in systems of interest, providing an explanation for why anti-parallel collisional reconnection shows effectively no spreading \citep{Nakamura12}. 
We confirm the validity of each of these analytical results using 3D two-fluid and resistive MHD numerical simulations. 
%We confirm this prediction with resistive MHD simulations and also show spreading may occur for resistive reconnection if a guide field is included in the initial conditions, consistent with our model. 

Our simulation study also reveals two other important aspects of reconnection spreading. First, we find that reconnection with current sheets wider than the ion inertial scale spread (for current sheets up to thicknesses of $3 d_i$), which differs from prior work in which reconnection convected with a fixed extent without spreading \citep{Shay03,Mayer13}. We hypothesize that the reason for the difference is that the prior work employed a computational domain with a square reconnecting geometry, while ours employ a rectangular geometry. The difference is likely caused by the additional free magnetic energy in the rectangular domain. Assessing this more carefully should be the subject of future research.

%{\bf Paragraph about possible understanding of spreading in the ion direction in Mike's simulations???}

Second, our simulation of anti-parallel reconnection with a perturbation in the plasma pressure to force reconnection instead of using a magnetic perturbation still results in reconnection spreading with the direction and speed of the current carriers. It does not spread at the faster magnetosonic speed, as has been previously postulated \citep{Vorpahl76}. Future research with other simulation setups are needed to more thoroughly test this prediction.

The results of the present study has a number of important implications. The analysis carried out here makes it clear that spreading with and without a guide field are dominated by different physics. Consequently, the MHD description is sufficient to describe the speed and bidirectionality of the spreading in reconnection with a strong guide field.  In contrast, the MHD description does not properly describe the spreading of anti-parallel reconnection, at least for the uniform resistivity profile employed in the present study. This has important implications for global 3D modeling in coronal and magnetospheric settings. Global MHD simulations have historically been quite common in both settings. Our results suggest that care is needed when the reconnection is anti-parallel in such settings.

Another implication is that the present research provides a formalism to understand reconnection in more general settings than studied here.  For example, reconnection spreading studies have been carried out for current sheets of uniform thickness. However, this is unlikely to be the case in many applications, such as the propagation of reconnection between two solar active regions in sympathetic flares, or in Earth's magnetotail or dayside magnetopause. The theory here can be used to understand spreading in such systems, as will be the topic of a forthcoming study (Arencibia et al., in preparation).

The theory presented here assumes reconnection spreads solely by inducing the x-line topology in  non-reconnecting regions. An alternate model is that a pressure minimum in the reconnecting region is convected into the non-reconnecting region, generating an inflow to initiate reconnection and thus elongate the x-line \citep{Huba03,Nakamura12,Dorfman13}. We note this model would predict that anti-parallel reconnection in resistive MHD would convect at the speed of the current carriers. However, numerical simulations in this work and in Nakamura et al.~(2012)\citep{Nakamura12} show no evidence of this. This suggests that induction of the reconnecting magnetic field in the out of plane direction, not the pressure gradient between reconnecting and non-reconnecting regions, is the essential ingredient for spreading. 

Our result may also be important in the context of recent results \citep{Zou18,tak} that the spreading of asymmetric guide field reconnection at Earth's dayside magnetopause occurs at Alfv\'enic speeds bidirectionally for thin current sheets, but at the slower speed of the current carriers for thicker current sheets. They argued that the reason is that the tearing instability is slower in thicker current sheets, and showed this hypothesis organizes when the spreading is Alfv\'enic or dominated by current carriers. Then, information about the onset time [alternately the growth time of the linear tearing mode \citep{tak}] is needed to find the time until reconnection onsets. If too slow, other spreading mechanisms could be faster. In particular, the present analysis suggests that different spreading mechanisms are additive, so convection of the reconnected magnetic field by the current carriers remains an active effect due to the Hall term in the generalized Ohm's law even if the convection term that drives spreading in thin current sheets is small for thicker current sheets. More research is needed to understand the interplay between the multiple possible spreading mechanisms.
%We hypothesize that the convection of a low pressure region into the non-reconnecting region, which drives an inflow which starts reconnection \citep{Huba03,Nakamura12}, could be such a mechanism. 

There are many avenues for future studies. For applications to the corona or magnetosphere, simulations in a 3D box may not faithfully capture the geometry of the system, including plasma and magnetic field asymmetries at the dayside magnetopause \citep{tak}, curvature of the large-scale magnetic fields, and density stratification in the corona.  As has been previously pointed out \citep{Qiu17}, some two-ribbon flare spreading events with nearly anti-parallel magnetic fields spread in the direction opposite to that of the inferred current carriers. This may be related to the global magnetic field configuration not captured in the treatment here. More work is needed to understand spreading in fundamentally 3D reconnection geometries \citep{Lukin11}.  Also, as discussed earlier, it is important to study the relationship between reconnection regions spreading vs.~convecting as a function of system parameters, including the computational domain size, the perturbation size and wavenumber \citep{Jain13}, the current sheet thickness, and when the gradient in the thickness of the current sheet is very large \citep{Huang19}.  Further work is also needed to understand how the spreading is impacted by the mechanism that reconnection is seeded, especially with a guide field. As anomalous resistivity has been invoked to explain reconnection in the solar corona \citep{Ugai77, Sato79}, it is also important to study spreading in with this dissipation mechanism.  
%or where spreading occurs in the direction of decreasing current sheet thickness.

\acknowledgments

Support is gratefully acknowledged from NSF Grants AGS-1460037, AGS-1602769, and PHY-1804428, NASA Grant NNX16AG76G, DOE Grant DE-SC0020294 (PAC) and NASA Grants NNX17AI25G, 80NSSC20K0198 and 80NSSC20K1813 (MAS). We acknowledge helpful conversations with M. Hasan Barbhuiya, James Drake, Heli Hietala, Haoming Liang, Dana Longcope, Tai Phan, Jiong Qiu, Kathy Reeves, and Luke Shepherd. Computational resources supporting this work were provided by the NASA High-End Computing (HEC) Program through the NASA Advanced Supercomputing (NAS) Division at Ames Research Center and by the National Energy Research Scientific Computing Center (NERSC), a DOE Office of Science User Facility supported by the Office of Science of the U.S.~Department of Energy under Contract No.~DE-AC02-05CH11231. The data that support the findings of this study are available from the corresponding authors upon reasonable request.

%\facility{facility ID}
%\facilities{NERSC} 
%\software{Numpy}

%\bibliography{gcrbib.bib}

\begin{thebibliography}{61}%
\makeatletter
\providecommand \@ifxundefined [1]{%
 \@ifx{#1\undefined}
}%
\providecommand \@ifnum [1]{%
 \ifnum #1\expandafter \@firstoftwo
 \else \expandafter \@secondoftwo
 \fi
}%
\providecommand \@ifx [1]{%
 \ifx #1\expandafter \@firstoftwo
 \else \expandafter \@secondoftwo
 \fi
}%
\providecommand \natexlab [1]{#1}%
\providecommand \enquote  [1]{``#1''}%
\providecommand \bibnamefont  [1]{#1}%
\providecommand \bibfnamefont [1]{#1}%
\providecommand \citenamefont [1]{#1}%
\providecommand \href@noop [0]{\@secondoftwo}%
\providecommand \href [0]{\begingroup \@sanitize@url \@href}%
\providecommand \@href[1]{\@@startlink{#1}\@@href}%
\providecommand \@@href[1]{\endgroup#1\@@endlink}%
\providecommand \@sanitize@url [0]{\catcode `\\12\catcode `\$12\catcode
  `\&12\catcode `\#12\catcode `\^12\catcode `\_12\catcode `\%12\relax}%
\providecommand \@@startlink[1]{}%
\providecommand \@@endlink[0]{}%
\providecommand \url  [0]{\begingroup\@sanitize@url \@url }%
\providecommand \@url [1]{\endgroup\@href {#1}{\urlprefix }}%
\providecommand \urlprefix  [0]{URL }%
\providecommand \Eprint [0]{\href }%
\providecommand \doibase [0]{http://dx.doi.org/}%
\providecommand \selectlanguage [0]{\@gobble}%
\providecommand \bibinfo  [0]{\@secondoftwo}%
\providecommand \bibfield  [0]{\@secondoftwo}%
\providecommand \translation [1]{[#1]}%
\providecommand \BibitemOpen [0]{}%
\providecommand \bibitemStop [0]{}%
\providecommand \bibitemNoStop [0]{.\EOS\space}%
\providecommand \EOS [0]{\spacefactor3000\relax}%
\providecommand \BibitemShut  [1]{\csname bibitem#1\endcsname}%
\let\auto@bib@innerbib\@empty
%</preamble>
\bibitem [{\citenamefont {Dungey}(1953)}]{Dungey53}%
  \BibitemOpen
  \bibfield  {author} {\bibinfo {author} {\bibfnamefont {J.~W.}\ \bibnamefont
  {Dungey}},\ }\bibfield  {title} {\enquote {\bibinfo {title} {Conditions for
  the occurrence of electrical discharges in astrophysical systems},}\
  }\href@noop {} {\bibfield  {journal} {\bibinfo  {journal} {Phil.~Mag.}\
  }\textbf {\bibinfo {volume} {44}},\ \bibinfo {pages} {725} (\bibinfo {year}
  {1953})}\BibitemShut {NoStop}%
\bibitem [{\citenamefont {Vasyliunas}(1975)}]{Vasyliunas75}%
  \BibitemOpen
  \bibfield  {author} {\bibinfo {author} {\bibfnamefont {V.~M.}\ \bibnamefont
  {Vasyliunas}},\ }\bibfield  {title} {\enquote {\bibinfo {title} {Theoretical
  models of magnetic field line merging, 1},}\ }\href@noop {} {\bibfield
  {journal} {\bibinfo  {journal} {Rev. Geophys.}\ }\textbf {\bibinfo {volume}
  {13}},\ \bibinfo {pages} {303} (\bibinfo {year} {1975})}\BibitemShut
  {NoStop}%
\bibitem [{\citenamefont {Priest}\ and\ \citenamefont
  {Forbes}(2000)}]{Priest00}%
  \BibitemOpen
  \bibfield  {author} {\bibinfo {author} {\bibfnamefont {E.}~\bibnamefont
  {Priest}}\ and\ \bibinfo {author} {\bibfnamefont {T.}~\bibnamefont
  {Forbes}},\ }\href@noop {} {\emph {\bibinfo {title} {Magnetic
  Reconnection}}}\ (\bibinfo  {publisher} {Cambridge University Press},\
  \bibinfo {year} {2000})\BibitemShut {NoStop}%
\bibitem [{\citenamefont {McPherron}, \citenamefont {Russell},\ and\
  \citenamefont {Aubry}(1973)}]{McPherron73}%
  \BibitemOpen
  \bibfield  {author} {\bibinfo {author} {\bibfnamefont {R.~L.}\ \bibnamefont
  {McPherron}}, \bibinfo {author} {\bibfnamefont {C.~T.}\ \bibnamefont
  {Russell}}, \ and\ \bibinfo {author} {\bibfnamefont {M.~P.}\ \bibnamefont
  {Aubry}},\ }\bibfield  {title} {\enquote {\bibinfo {title} {Phenomenological
  model for substorms},}\ }\href@noop {} {\bibfield  {journal} {\bibinfo
  {journal} {J. Geophys. Res.}\ }\textbf {\bibinfo {volume} {78}},\ \bibinfo
  {pages} {3131} (\bibinfo {year} {1973})}\BibitemShut {NoStop}%
\bibitem [{\citenamefont {{Uzdensky}}(2011)}]{Uzdenski}%
  \BibitemOpen
  \bibfield  {author} {\bibinfo {author} {\bibfnamefont {D.~A.}\ \bibnamefont
  {{Uzdensky}}},\ }\bibfield  {title} {\enquote {\bibinfo {title} {{Magnetic
  Reconnection in Extreme Astrophysical Environments}},}\ }\href {\doibase
  10.1007/s11214-011-9744-5} {\bibfield  {journal} {\bibinfo  {journal} {Space Sci.~Rev.} \
  }\textbf {\bibinfo {volume} {160}},\ \bibinfo {pages} {45--71} (\bibinfo
  {year} {2011})},\ \Eprint {http://arxiv.org/abs/1101.2472} {arXiv:1101.2472
  [astro-ph.HE]} \BibitemShut {NoStop}%
\bibitem [{\citenamefont {Uzdensky}(2016)}]{Uzdensky16}%
  \BibitemOpen
  \bibfield  {author} {\bibinfo {author} {\bibfnamefont {D.~A.}\ \bibnamefont
  {Uzdensky}},\ }\bibfield  {title} {\enquote {\bibinfo {title} {Radiative
  {Magnetic} {Reconnection} in {Astrophysics}},}\ }in\ \href {\doibase
  10.1007/978-3-319-26432-5_12} {\emph {\bibinfo {booktitle} {Magnetic
  {Reconnection}: {Concepts} and {Applications}}}},\ \bibinfo {series and
  number} {Astrophysics and {Space} {Science} {Library}},\ \bibinfo {editor}
  {edited by\ \bibinfo {editor} {\bibfnamefont {W.}~\bibnamefont {Gonzalez}}\
  and\ \bibinfo {editor} {\bibfnamefont {E.}~\bibnamefont {Parker}}}\ (\bibinfo
   {publisher} {Springer International Publishing},\ \bibinfo {address}
  {Cham},\ \bibinfo {year} {2016})\ pp.\ \bibinfo {pages}
  {473--519}\BibitemShut {NoStop}%
\bibitem [{\citenamefont {Sweet}(1958)}]{Sweet58}%
  \BibitemOpen
  \bibfield  {author} {\bibinfo {author} {\bibfnamefont {P.~A.}\ \bibnamefont
  {Sweet}},\ }\bibfield  {title} {\enquote {\bibinfo {title} {The neutral point
  theory of solar flares},}\ }in\ \href@noop {} {\emph {\bibinfo {booktitle}
  {Electromagnetic Phenomena in Cosmical Physics}}},\ \bibinfo {editor} {edited
  by\ \bibinfo {editor} {\bibfnamefont {B.}~\bibnamefont {Lehnert}}}\ (\bibinfo
   {publisher} {Cambridge University Press, New York},\ \bibinfo {year}
  {1958})\ p.\ \bibinfo {pages} {123}\BibitemShut {NoStop}%
\bibitem [{\citenamefont {Parker}(1957)}]{Parker57}%
  \BibitemOpen
  \bibfield  {author} {\bibinfo {author} {\bibfnamefont {E.~N.}\ \bibnamefont
  {Parker}},\ }\bibfield  {title} {\enquote {\bibinfo {title} {Sweet's
  mechanism for merging magnetic fields in conducting fluids},}\ }\href@noop {}
  {\bibfield  {journal} {\bibinfo  {journal} {J. Geophys. Res.}\ }\textbf
  {\bibinfo {volume} {62}},\ \bibinfo {pages} {509} (\bibinfo {year}
  {1957})}\BibitemShut {NoStop}%
\bibitem [{\citenamefont {Petschek}(1964)}]{petschek64a}%
  \BibitemOpen
  \bibfield  {author} {\bibinfo {author} {\bibfnamefont {H.~E.}\ \bibnamefont
  {Petschek}},\ }\bibfield  {title} {\enquote {\bibinfo {title} {Magnetic field
  annihilation},}\ }in\ \href@noop {} {\emph {\bibinfo {booktitle} {Proc.
  AAS-NASA Symp. Phys. Solar Flares}}},\ \bibinfo {series} {NASA-SP},
  Vol.~\bibinfo {volume} {50}\ (\bibinfo {year} {1964})\ pp.\ \bibinfo {pages}
  {425--439}\BibitemShut {NoStop}%
\bibitem [{\citenamefont {Pontin}(2011)}]{Pontin11}%
  \BibitemOpen
  \bibfield  {author} {\bibinfo {author} {\bibfnamefont {D.~I.}\ \bibnamefont
  {Pontin}},\ }\bibfield  {title} {\enquote {\bibinfo {title}
  {Three-dimensional magnetic reconnection regimes: A review},}\ }\href@noop {}
  {\bibfield  {journal} {\bibinfo  {journal} {Adv.~Space Res.}\ }\textbf
  {\bibinfo {volume} {47}},\ \bibinfo {pages} {1508} (\bibinfo {year}
  {2011})}\BibitemShut {NoStop}%
\bibitem [{\citenamefont {Lukin}\ and\ \citenamefont {Linton}(2011)}]{Lukin11}%
  \BibitemOpen
  \bibfield  {author} {\bibinfo {author} {\bibfnamefont {V.~S.}\ \bibnamefont
  {Lukin}}\ and\ \bibinfo {author} {\bibfnamefont {M.~G.}\ \bibnamefont
  {Linton}},\ }\bibfield  {title} {\enquote {\bibinfo {title}
  {Three-dimensional magnetic reconnection through a moving magnetic null},}\
  }\href@noop {} {\bibfield  {journal} {\bibinfo  {journal} {Nonlinear
  Processes in Geophysics}\ }\textbf {\bibinfo {volume} {18}},\ \bibinfo
  {pages} {871} (\bibinfo {year} {2011})}\BibitemShut {NoStop}%
\bibitem [{\citenamefont {Priest}(2014)}]{Priest14}%
  \BibitemOpen
  \bibfield  {author} {\bibinfo {author} {\bibfnamefont {E.}~\bibnamefont
  {Priest}},\ }\href@noop {} {\emph {\bibinfo {title} {Magnetohydrodynamics of
  the Sun}}}\ (\bibinfo  {publisher} {Cambridge University Press},\ \bibinfo
  {address} {New York},\ \bibinfo {year} {2014})\BibitemShut {NoStop}%
\bibitem [{\citenamefont {Shay}\ \emph {et~al.}(2003)\citenamefont {Shay},
  \citenamefont {Drake}, \citenamefont {Swisdak}, \citenamefont {Dorland},\
  and\ \citenamefont {Rogers}}]{Shay03}%
  \BibitemOpen
  \bibfield  {author} {\bibinfo {author} {\bibfnamefont {M.~A.}\ \bibnamefont
  {Shay}}, \bibinfo {author} {\bibfnamefont {J.~F.}\ \bibnamefont {Drake}},
  \bibinfo {author} {\bibfnamefont {M.}~\bibnamefont {Swisdak}}, \bibinfo
  {author} {\bibfnamefont {W.}~\bibnamefont {Dorland}}, \ and\ \bibinfo
  {author} {\bibfnamefont {B.~N.}\ \bibnamefont {Rogers}},\ }\bibfield  {title}
  {\enquote {\bibinfo {title} {Inherently three-dimensional magnetic
  reconnection: {A} mechanism for bursty bulk flows?}}\ }\href@noop {}
  {\bibfield  {journal} {\bibinfo  {journal} {Geophys. Res. Lett.}\ }\textbf
  {\bibinfo {volume} {30}},\ \bibinfo {pages} {1345} (\bibinfo {year}
  {2003})}\BibitemShut {NoStop}%
\bibitem [{\citenamefont {Linton}\ and\ \citenamefont
  {Longcope}(2006)}]{Linton06}%
  \BibitemOpen
  \bibfield  {author} {\bibinfo {author} {\bibfnamefont {M.~G.}\ \bibnamefont
  {Linton}}\ and\ \bibinfo {author} {\bibfnamefont {D.~W.}\ \bibnamefont
  {Longcope}},\ }\bibfield  {title} {\enquote {\bibinfo {title} {A model for
  patchy reconnection in three dimensions},}\ }\href@noop {} {\bibfield
  {journal} {\bibinfo  {journal} {Ap.~J.}\ }\textbf {\bibinfo {volume} {642}},\
  \bibinfo {pages} {1177} (\bibinfo {year} {2006})}\BibitemShut {NoStop}%
\bibitem [{\citenamefont {{Meyer III}}(2013)}]{Mayer13}%
  \BibitemOpen
  \bibfield  {author} {\bibinfo {author} {\bibfnamefont {J.~C.}\ \bibnamefont
  {{Meyer III}}},\ }\emph {\bibinfo {title} {Structure of the Diffusion Region
  in Three Dimensional Magnetic Reconnection}},\ \href@noop {} {Ph.D. thesis},\
  \bibinfo  {school} {University of Delaware} (\bibinfo {year}
  {2013})\BibitemShut {NoStop}%
\bibitem [{\citenamefont {{Sasunov}}\ \emph {et~al.}(2015)\citenamefont
  {{Sasunov}}, \citenamefont {{Semenov}}, \citenamefont {{Heyn}}, \citenamefont
  {{Erkaev}}, \citenamefont {{Kubyshkin}}, \citenamefont {{Slivka}},
  \citenamefont {{Korovinskiy}},\ and\ \citenamefont
  {{Khodachenko}}}]{Sasunov2015}%
  \BibitemOpen
  \bibfield  {author} {\bibinfo {author} {\bibfnamefont {Y.~L.}\ \bibnamefont
  {{Sasunov}}}, \bibinfo {author} {\bibfnamefont {V.~S.}\ \bibnamefont
  {{Semenov}}}, \bibinfo {author} {\bibfnamefont {M.~F.}\ \bibnamefont
  {{Heyn}}}, \bibinfo {author} {\bibfnamefont {N.~V.}\ \bibnamefont
  {{Erkaev}}}, \bibinfo {author} {\bibfnamefont {I.~V.}\ \bibnamefont
  {{Kubyshkin}}}, \bibinfo {author} {\bibfnamefont {K.~Y.}\ \bibnamefont
  {{Slivka}}}, \bibinfo {author} {\bibfnamefont {D.~B.}\ \bibnamefont
  {{Korovinskiy}}}, \ and\ \bibinfo {author} {\bibfnamefont {M.~L.}\
  \bibnamefont {{Khodachenko}}},\ }\bibfield  {title} {\enquote {\bibinfo
  {title} {{A statistical survey of reconnection exhausts in the solar wind
  based on the Riemannian decay of current sheets}},}\ }\href {\doibase
  10.1002/2015JA021504} {\bibfield  {journal} {\bibinfo  {journal} {Journal of
  Geophysical Research (Space Physics)}\ }\textbf {\bibinfo {volume} {120}},\
  \bibinfo {pages} {8194--8209} (\bibinfo {year} {2015})}\BibitemShut {NoStop}%
\bibitem [{\citenamefont {Shepherd}\ \emph {et~al.}(2017)\citenamefont
  {Shepherd}, \citenamefont {Cassak}, \citenamefont {Drake}, \citenamefont
  {Gosling}, \citenamefont {Phan},\ and\ \citenamefont {Shay}}]{Shepherd17}%
  \BibitemOpen
  \bibfield  {author} {\bibinfo {author} {\bibfnamefont {L.~S.}\ \bibnamefont
  {Shepherd}}, \bibinfo {author} {\bibfnamefont {P.~A.}\ \bibnamefont
  {Cassak}}, \bibinfo {author} {\bibfnamefont {J.~F.}\ \bibnamefont {Drake}},
  \bibinfo {author} {\bibfnamefont {J.~T.}\ \bibnamefont {Gosling}}, \bibinfo
  {author} {\bibfnamefont {T.-D.}\ \bibnamefont {Phan}}, \ and\ \bibinfo
  {author} {\bibfnamefont {M.~A.}\ \bibnamefont {Shay}},\ }\bibfield  {title}
  {\enquote {\bibinfo {title} {Structure of exhausts in magnetic reconnection
  with an x-line of finite extent},}\ }\href@noop {} {\bibfield  {journal}
  {\bibinfo  {journal} {Ap.~J.}\ }\textbf {\bibinfo {volume} {848}},\ \bibinfo
  {pages} {90} (\bibinfo {year} {2017})}\BibitemShut {NoStop}%
\bibitem [{\citenamefont {Liu}\ \emph {et~al.}(2019)\citenamefont {Liu},
  \citenamefont {Li}, \citenamefont {Hesse}, \citenamefont {Sun}, \citenamefont
  {Liu}, \citenamefont {Burch}, \citenamefont {Slavin},\ and\ \citenamefont
  {Huang}}]{Liu19}%
  \BibitemOpen
  \bibfield  {author} {\bibinfo {author} {\bibfnamefont {Y.-H.}\ \bibnamefont
  {Liu}}, \bibinfo {author} {\bibfnamefont {T.~C.}\ \bibnamefont {Li}},
  \bibinfo {author} {\bibfnamefont {M.}~\bibnamefont {Hesse}}, \bibinfo
  {author} {\bibfnamefont {W.~J.}\ \bibnamefont {Sun}}, \bibinfo {author}
  {\bibfnamefont {J.}~\bibnamefont {Liu}}, \bibinfo {author} {\bibfnamefont
  {J.}~\bibnamefont {Burch}}, \bibinfo {author} {\bibfnamefont {J.~A.}\
  \bibnamefont {Slavin}}, \ and\ \bibinfo {author} {\bibfnamefont
  {K.}~\bibnamefont {Huang}},\ }\bibfield  {title} {\enquote {\bibinfo {title}
  {Three-{Dimensional} {Magnetic} {Reconnection} {With} a {Spatially}
  {Confined} {X}-{Line} {Extent}: {Implications} for {Dipolarizing} {Flux}
  {Bundles} and the {Dawn}-{Dusk} {Asymmetry}},}\ }\href {\doibase
  https://doi.org/10.1029/2019JA026539} {\bibfield  {journal} {\bibinfo
  {journal} {Journal of Geophysical Research: Space Physics}\ }\textbf
  {\bibinfo {volume} {124}},\ \bibinfo {pages} {2819--2830} (\bibinfo {year}
  {2019})}\BibitemShut {NoStop}%
\bibitem [{\citenamefont {Huang}\ \emph {et~al.}(2020)\citenamefont {Huang},
  \citenamefont {Liu}, \citenamefont {Lu},\ and\ \citenamefont
  {Hesse}}]{Huang19}%
  \BibitemOpen
  \bibfield  {author} {\bibinfo {author} {\bibfnamefont {K.}~\bibnamefont
  {Huang}}, \bibinfo {author} {\bibfnamefont {Y.-H.}\ \bibnamefont {Liu}},
  \bibinfo {author} {\bibfnamefont {Q.}~\bibnamefont {Lu}}, \ and\ \bibinfo
  {author} {\bibfnamefont {M.}~\bibnamefont {Hesse}},\ }\bibfield  {title}
  {\enquote {\bibinfo {title} {Scaling of magnetic reconnection with a limited
  x-line extent},}\ }\href {\doibase 10.1029/2020GL088147} {\bibfield
  {journal} {\bibinfo  {journal} {Geophysical Research Letters}\ }\textbf
  {\bibinfo {volume} {47}},\ \bibinfo {pages} {e2020GL088147} (\bibinfo {year}
  {2020})}\BibitemShut {NoStop}%
\bibitem [{\citenamefont {Isobe}\ \emph {et~al.}(2002)\citenamefont {Isobe},
  \citenamefont {Yokoyama}, \citenamefont {Shimojo}, \citenamefont {Morimoto},
  \citenamefont {Kozu}, \citenamefont {Eto}, \citenamefont {Narukage},\ and\
  \citenamefont {Shibata}}]{Isobe02}%
  \BibitemOpen
  \bibfield  {author} {\bibinfo {author} {\bibfnamefont {H.}~\bibnamefont
  {Isobe}}, \bibinfo {author} {\bibfnamefont {T.}~\bibnamefont {Yokoyama}},
  \bibinfo {author} {\bibfnamefont {M.}~\bibnamefont {Shimojo}}, \bibinfo
  {author} {\bibfnamefont {T.}~\bibnamefont {Morimoto}}, \bibinfo {author}
  {\bibfnamefont {H.}~\bibnamefont {Kozu}}, \bibinfo {author} {\bibfnamefont
  {S.}~\bibnamefont {Eto}}, \bibinfo {author} {\bibfnamefont {N.}~\bibnamefont
  {Narukage}}, \ and\ \bibinfo {author} {\bibfnamefont {K.}~\bibnamefont
  {Shibata}},\ }\bibfield  {title} {\enquote {\bibinfo {title} {Reconnection
  rate in the decay phase of a long duration event flare on 1997 {M}ay 12},}\
  }\href@noop {} {\bibfield  {journal} {\bibinfo  {journal} {Ap.~J.}\ }\textbf
  {\bibinfo {volume} {566}},\ \bibinfo {pages} {528} (\bibinfo {year}
  {2002})}\BibitemShut {NoStop}%
\bibitem [{\citenamefont {Lee}\ and\ \citenamefont {Gary}(2008)}]{Lee08}%
  \BibitemOpen
  \bibfield  {author} {\bibinfo {author} {\bibfnamefont {J.}~\bibnamefont
  {Lee}}\ and\ \bibinfo {author} {\bibfnamefont {D.~E.}\ \bibnamefont {Gary}},\
  }\bibfield  {title} {\enquote {\bibinfo {title} {Parallel {Motions} of
  {Coronal} {Hard} {X}-{Ray} {Source} and {H $\alpha$} {Ribbons}},}\ }\href {\doibase
  10.1086/592292} {\bibfield  {journal} {\bibinfo  {journal} {ApJ}\ }\textbf
  {\bibinfo {volume} {685}},\ \bibinfo {pages} {L87} (\bibinfo {year}
  {2008})}\BibitemShut {NoStop}%
\bibitem [{\citenamefont {Qiu}(2009)}]{Qiu09}%
  \BibitemOpen
  \bibfield  {author} {\bibinfo {author} {\bibfnamefont {J.}~\bibnamefont
  {Qiu}},\ }\bibfield  {title} {\enquote {\bibinfo {title} {Observational
  analysis of magnetic reconnection sequence},}\ }\href@noop {} {\bibfield
  {journal} {\bibinfo  {journal} {Ap.~J.}\ }\textbf {\bibinfo {volume} {692}},\
  \bibinfo {pages} {1110} (\bibinfo {year} {2009})}\BibitemShut {NoStop}%
\bibitem [{\citenamefont {Qiu}\ \emph {et~al.}(2010)\citenamefont {Qiu},
  \citenamefont {Liu}, \citenamefont {Hill},\ and\ \citenamefont
  {Kazachenko}}]{Qiu10}%
  \BibitemOpen
  \bibfield  {author} {\bibinfo {author} {\bibfnamefont {J.}~\bibnamefont
  {Qiu}}, \bibinfo {author} {\bibfnamefont {W.}~\bibnamefont {Liu}}, \bibinfo
  {author} {\bibfnamefont {N.}~\bibnamefont {Hill}}, \ and\ \bibinfo {author}
  {\bibfnamefont {M.}~\bibnamefont {Kazachenko}},\ }\bibfield  {title}
  {\enquote {\bibinfo {title} {Reconnection and energetics in two-ribbon
  flares: A revisit of the {B}astille {D}ay flare},}\ }\href@noop {} {\bibfield
   {journal} {\bibinfo  {journal} {Ap.~J.}\ }\textbf {\bibinfo {volume}
  {725}},\ \bibinfo {pages} {319} (\bibinfo {year} {2010})}\BibitemShut
  {NoStop}%
\bibitem [{\citenamefont {Liu}\ \emph {et~al.}(2010)\citenamefont {Liu},
  \citenamefont {Lee}, \citenamefont {Jing}, \citenamefont {Liu}, \citenamefont
  {Deng},\ and\ \citenamefont {Wang}}]{Liu10ribbons}%
  \BibitemOpen
  \bibfield  {author} {\bibinfo {author} {\bibfnamefont {C.}~\bibnamefont
  {Liu}}, \bibinfo {author} {\bibfnamefont {J.}~\bibnamefont {Lee}}, \bibinfo
  {author} {\bibfnamefont {J.}~\bibnamefont {Jing}}, \bibinfo {author}
  {\bibfnamefont {R.}~\bibnamefont {Liu}}, \bibinfo {author} {\bibfnamefont
  {N.}~\bibnamefont {Deng}}, \ and\ \bibinfo {author} {\bibfnamefont
  {H.}~\bibnamefont {Wang}},\ }\bibfield  {title} {\enquote {\bibinfo {title}
  {{MOTIONS} {OF} {HARD} {X}-{RAY} {SOURCES} {DURING} {AN} {ASYMMETRIC}
  {ERUPTION}},}\ }\href {\doibase 10.1088/2041-8205/721/2/L193} {\bibfield
  {journal} {\bibinfo  {journal} {ApJL}\ }\textbf {\bibinfo {volume} {721}},\
  \bibinfo {pages} {L193--L198} (\bibinfo {year} {2010})}\BibitemShut {NoStop}%
\bibitem [{\citenamefont {Cheng}, \citenamefont {Kerr},\ and\ \citenamefont
  {Qiu}(2011)}]{Cheng11}%
  \BibitemOpen
  \bibfield  {author} {\bibinfo {author} {\bibfnamefont {J.~X.}\ \bibnamefont
  {Cheng}}, \bibinfo {author} {\bibfnamefont {G.}~\bibnamefont {Kerr}}, \ and\
  \bibinfo {author} {\bibfnamefont {J.}~\bibnamefont {Qiu}},\ }\bibfield
  {title} {\enquote {\bibinfo {title} {{HARD} {X}-{RAY} {AND} {ULTRAVIOLET}
  {OBSERVATIONS} {OF} {THE} 2005 {JANUARY} 15 {TWO}-{RIBBON} {FLARE}},}\ }\href
  {\doibase 10.1088/0004-637X/744/1/48} {\bibfield  {journal} {\bibinfo
  {journal} {ApJ}\ }\textbf {\bibinfo {volume} {744}},\ \bibinfo {pages} {48}
  (\bibinfo {year} {2011})}\BibitemShut {NoStop}%
\bibitem [{\citenamefont {{Tian}}\ \emph {et~al.}(2015)\citenamefont {{Tian}},
  \citenamefont {{Young}}, \citenamefont {{Reeves}}, \citenamefont {{Chen}},
  \citenamefont {{Liu}},\ and\ \citenamefont {{McKillop}}}]{Tian15}%
  \BibitemOpen
  \bibfield  {author} {\bibinfo {author} {\bibfnamefont {H.}~\bibnamefont
  {{Tian}}}, \bibinfo {author} {\bibfnamefont {P.~R.}\ \bibnamefont {{Young}}},
  \bibinfo {author} {\bibfnamefont {K.~K.}\ \bibnamefont {{Reeves}}}, \bibinfo
  {author} {\bibfnamefont {B.}~\bibnamefont {{Chen}}}, \bibinfo {author}
  {\bibfnamefont {W.}~\bibnamefont {{Liu}}}, \ and\ \bibinfo {author}
  {\bibfnamefont {S.}~\bibnamefont {{McKillop}}},\ }\bibfield  {title}
  {\enquote {\bibinfo {title} {{Temporal Evolution of Chromospheric
  Evaporation: Case Studies of the M1.1 Flare on 2014 September 6 and X1.6
  Flare on 2014 September 10}},}\ }\href {\doibase 10.1088/0004-637X/811/2/139}
  {\bibfield  {journal} {\bibinfo  {journal} {ApJ}\ }\textbf {\bibinfo
  {volume} {811}},\ \bibinfo {eid} {139} (\bibinfo {year} {2015})}\BibitemShut
  {NoStop}%
\bibitem [{\citenamefont {{Graham}}\ and\ \citenamefont
  {{Cauzzi}}(2015)}]{Graham15}%
  \BibitemOpen
  \bibfield  {author} {\bibinfo {author} {\bibfnamefont {D.~R.}\ \bibnamefont
  {{Graham}}}\ and\ \bibinfo {author} {\bibfnamefont {G.}~\bibnamefont
  {{Cauzzi}}},\ }\bibfield  {title} {\enquote {\bibinfo {title} {{Temporal
  Evolution of Multiple Evaporating Ribbon Sources in a Solar Flare}},}\ }\href
  {\doibase 10.1088/2041-8205/807/2/L22} {\bibfield  {journal} {\bibinfo
  {journal} {ApJL}\ }\textbf {\bibinfo {volume} {807}},\ \bibinfo {eid} {L22}
  (\bibinfo {year} {2015})}\BibitemShut {NoStop}%
\bibitem [{\citenamefont {{Qiu}}\ \emph {et~al.}(2017)\citenamefont {{Qiu}},
  \citenamefont {{Longcope}}, \citenamefont {{Cassak}},\ and\ \citenamefont
  {{Priest}}}]{Qiu17}%
  \BibitemOpen
  \bibfield  {author} {\bibinfo {author} {\bibfnamefont {J.}~\bibnamefont
  {{Qiu}}}, \bibinfo {author} {\bibfnamefont {D.~W.}\ \bibnamefont
  {{Longcope}}}, \bibinfo {author} {\bibfnamefont {P.~A.}\ \bibnamefont
  {{Cassak}}}, \ and\ \bibinfo {author} {\bibfnamefont {E.~R.}\ \bibnamefont
  {{Priest}}},\ }\bibfield  {title} {\enquote {\bibinfo {title} {{Elongation of
  Flare Ribbons}},}\ }\href {\doibase 10.3847/1538-4357/aa6341} {\bibfield
  {journal} {\bibinfo  {journal} {ApJ}\ }\textbf {\bibinfo {volume} {838}},\
  \bibinfo {eid} {17} (\bibinfo {year} {2017})}\BibitemShut {NoStop}%
\bibitem [{\citenamefont {Tripathi}, \citenamefont {Isobe},\ and\ \citenamefont
  {Mason}(2006)}]{Tripathi06}%
  \BibitemOpen
  \bibfield  {author} {\bibinfo {author} {\bibfnamefont {D.}~\bibnamefont
  {Tripathi}}, \bibinfo {author} {\bibfnamefont {H.}~\bibnamefont {Isobe}}, \
  and\ \bibinfo {author} {\bibfnamefont {H.~E.}\ \bibnamefont {Mason}},\
  }\bibfield  {title} {\enquote {\bibinfo {title} {On the propagation of
  brightening after filament/prominence eruptions, as seen by {SoHO-EIT}},}\
  }\href@noop {} {\bibfield  {journal} {\bibinfo  {journal}
  {Astron.~Astrophys.}\ }\textbf {\bibinfo {volume} {453}},\ \bibinfo {pages}
  {1111} (\bibinfo {year} {2006})}\BibitemShut {NoStop}%
\bibitem [{\citenamefont {{Zhou}}\ \emph {et~al.}(2017)\citenamefont {{Zhou}},
  \citenamefont {{Ashour-Abdalla}}, \citenamefont {{Deng}}, \citenamefont
  {{Pang}}, \citenamefont {{Fu}}, \citenamefont {{Walker}}, \citenamefont
  {{Lapenta}}, \citenamefont {{Huang}}, \citenamefont {{Xu}},\ and\
  \citenamefont {{Tang}}}]{Zhou17}%
  \BibitemOpen
  \bibfield  {author} {\bibinfo {author} {\bibfnamefont {M.}~\bibnamefont
  {{Zhou}}}, \bibinfo {author} {\bibfnamefont {M.}~\bibnamefont
  {{Ashour-Abdalla}}}, \bibinfo {author} {\bibfnamefont {X.}~\bibnamefont
  {{Deng}}}, \bibinfo {author} {\bibfnamefont {Y.}~\bibnamefont {{Pang}}},
  \bibinfo {author} {\bibfnamefont {H.}~\bibnamefont {{Fu}}}, \bibinfo {author}
  {\bibfnamefont {R.}~\bibnamefont {{Walker}}}, \bibinfo {author}
  {\bibfnamefont {G.}~\bibnamefont {{Lapenta}}}, \bibinfo {author}
  {\bibfnamefont {S.}~\bibnamefont {{Huang}}}, \bibinfo {author} {\bibfnamefont
  {X.}~\bibnamefont {{Xu}}}, \ and\ \bibinfo {author} {\bibfnamefont
  {R.}~\bibnamefont {{Tang}}},\ }\bibfield  {title} {\enquote {\bibinfo {title}
  {{Observation of Three-Dimensional Magnetic Reconnection in the Terrestrial
  Magnetotail}},}\ }\href {\doibase 10.1002/2017JA024597} {\bibfield  {journal}
  {\bibinfo  {journal} {Journal of Geophysical Research (Space Physics)}\
  }\textbf {\bibinfo {volume} {122}},\ \bibinfo {pages} {9513--9520} (\bibinfo
  {year} {2017})}\BibitemShut {NoStop}%
\bibitem [{\citenamefont {{Zou}}\ \emph {et~al.}(2018)\citenamefont {{Zou}},
  \citenamefont {{Walsh}}, \citenamefont {{Nishimura}}, \citenamefont
  {{Angelopoulos}}, \citenamefont {{Ruohoniemi}}, \citenamefont
  {{McWilliams}},\ and\ \citenamefont {{Nishitani}}}]{Zou18}%
  \BibitemOpen
  \bibfield  {author} {\bibinfo {author} {\bibfnamefont {Y.}~\bibnamefont
  {{Zou}}}, \bibinfo {author} {\bibfnamefont {B.~M.}\ \bibnamefont {{Walsh}}},
  \bibinfo {author} {\bibfnamefont {Y.}~\bibnamefont {{Nishimura}}}, \bibinfo
  {author} {\bibfnamefont {V.}~\bibnamefont {{Angelopoulos}}}, \bibinfo
  {author} {\bibfnamefont {J.~M.}\ \bibnamefont {{Ruohoniemi}}}, \bibinfo
  {author} {\bibfnamefont {K.~A.}\ \bibnamefont {{McWilliams}}}, \ and\
  \bibinfo {author} {\bibfnamefont {N.}~\bibnamefont {{Nishitani}}},\
  }\bibfield  {title} {\enquote {\bibinfo {title} {{Spreading Speed of
  Magnetopause Reconnection X-Lines Using Ground-Satellite Coordination}},}\
  }\href {\doibase 10.1002/2017GL075765} {\bibfield  {journal} {\bibinfo
  {journal} {Geophys.~Res.~Lett.}\ }\textbf {\bibinfo {volume} {45}},\ \bibinfo {pages}
  {80--89} (\bibinfo {year} {2018})}\BibitemShut {NoStop}%
\bibitem [{\citenamefont {Zou}\ \emph {et~al.}(2019)\citenamefont {Zou},
  \citenamefont {Walsh}, \citenamefont {Nishimura}, \citenamefont
  {Angelopoulos}, \citenamefont {Ruohoniemi}, \citenamefont {McWilliams},\ and\
  \citenamefont {Nishitani}}]{Zou20}%
  \BibitemOpen
  \bibfield  {author} {\bibinfo {author} {\bibfnamefont {Y.}~\bibnamefont
  {Zou}}, \bibinfo {author} {\bibfnamefont {B.~M.}\ \bibnamefont {Walsh}},
  \bibinfo {author} {\bibfnamefont {Y.}~\bibnamefont {Nishimura}}, \bibinfo
  {author} {\bibfnamefont {V.}~\bibnamefont {Angelopoulos}}, \bibinfo {author}
  {\bibfnamefont {J.~M.}\ \bibnamefont {Ruohoniemi}}, \bibinfo {author}
  {\bibfnamefont {K.~A.}\ \bibnamefont {McWilliams}}, \ and\ \bibinfo {author}
  {\bibfnamefont {N.}~\bibnamefont {Nishitani}},\ }\bibfield  {title}
  {{ {\bibinfo {title} {Local time extent of
  magnetopause reconnection using space–ground coordination},}\ }}\href
  {\doibase https://doi.org/10.5194/angeo-37-215-2019} {\bibfield  {journal}
  {\bibinfo  {journal} {Annales Geophysicae}\ }\textbf {\bibinfo {volume}
  {37}},\ \bibinfo {pages} {215--234} (\bibinfo {year} {2019})}\BibitemShut
  {NoStop}%
\bibitem [{\citenamefont {Nagai}(1982)}]{Nagai82}%
  \BibitemOpen
  \bibfield  {author} {\bibinfo {author} {\bibfnamefont {T.}~\bibnamefont
  {Nagai}},\ }\bibfield  {title} {\enquote {\bibinfo {title} {Observed magnetic
  substorm signatures at synchronous altitude},}\ }\href@noop {} {\bibfield
  {journal} {\bibinfo  {journal} {J. Geophys. Res.}\ }\textbf {\bibinfo
  {volume} {87}},\ \bibinfo {pages} {4405} (\bibinfo {year}
  {1982})}\BibitemShut {NoStop}%
\bibitem [{\citenamefont {Nagai}\ \emph {et~al.}(2013)\citenamefont {Nagai},
  \citenamefont {Shinohara}, \citenamefont {Zenitani}, \citenamefont
  {Nakamura}, \citenamefont {Nakamura}, \citenamefont {Fujimoto}, \citenamefont
  {Saito},\ and\ \citenamefont {Mukai}}]{Nagai13}%
  \BibitemOpen
  \bibfield  {author} {\bibinfo {author} {\bibfnamefont {T.}~\bibnamefont
  {Nagai}}, \bibinfo {author} {\bibfnamefont {I.}~\bibnamefont {Shinohara}},
  \bibinfo {author} {\bibfnamefont {S.}~\bibnamefont {Zenitani}}, \bibinfo
  {author} {\bibfnamefont {R.}~\bibnamefont {Nakamura}}, \bibinfo {author}
  {\bibfnamefont {T.~K.~M.}\ \bibnamefont {Nakamura}}, \bibinfo {author}
  {\bibfnamefont {M.}~\bibnamefont {Fujimoto}}, \bibinfo {author}
  {\bibfnamefont {Y.}~\bibnamefont {Saito}}, \ and\ \bibinfo {author}
  {\bibfnamefont {T.}~\bibnamefont {Mukai}},\ }\bibfield  {title} {\enquote
  {\bibinfo {title} {Three-dimensional structure of magnetic reconnection in
  the magnetotail from geotail observations},}\ }\href {\doibase
  10.1002/jgra.50247} {\bibfield  {journal} {\bibinfo  {journal} {Journal of
  Geophysical Research: Space Physics}\ }\textbf {\bibinfo {volume} {118}},\
  \bibinfo {pages} {1667--1678} (\bibinfo {year} {2013})}\BibitemShut {NoStop}%
\bibitem [{\citenamefont {Hietala}, \citenamefont {Eastwood},\ and\
  \citenamefont {Isavnin}(2014)}]{Hietala14}%
  \BibitemOpen
  \bibfield  {author} {\bibinfo {author} {\bibfnamefont {H.}~\bibnamefont
  {Hietala}}, \bibinfo {author} {\bibfnamefont {J.~P.}\ \bibnamefont
  {Eastwood}}, \ and\ \bibinfo {author} {\bibfnamefont {A.}~\bibnamefont
  {Isavnin}},\ }\bibfield  {title} {\enquote {\bibinfo {title} {Sequentially
  released tilted flux ropes in the earth’s magnetotail},}\ }\href@noop {}
  {\bibfield  {journal} {\bibinfo  {journal} {Plasma Phys. Control. Fusion}\
  }\textbf {\bibinfo {volume} {56}},\ \bibinfo {pages} {064011} (\bibinfo
  {year} {2014})}\BibitemShut {NoStop}%
\bibitem [{\citenamefont {Katz}\ \emph {et~al.}(2010)\citenamefont {Katz},
  \citenamefont {Egedal}, \citenamefont {Fox}, \citenamefont {Le},
  \citenamefont {Bonde},\ and\ \citenamefont {Vrublevskis}}]{Katz10}%
  \BibitemOpen
  \bibfield  {author} {\bibinfo {author} {\bibfnamefont {N.}~\bibnamefont
  {Katz}}, \bibinfo {author} {\bibfnamefont {J.}~\bibnamefont {Egedal}},
  \bibinfo {author} {\bibfnamefont {W.}~\bibnamefont {Fox}}, \bibinfo {author}
  {\bibfnamefont {A.}~\bibnamefont {Le}}, \bibinfo {author} {\bibfnamefont
  {J.}~\bibnamefont {Bonde}}, \ and\ \bibinfo {author} {\bibfnamefont
  {A.}~\bibnamefont {Vrublevskis}},\ }\bibfield  {title} {\enquote {\bibinfo
  {title} {Laboratory observation of localized onset of magnetic
  reconnection},}\ }\href@noop {} {\bibfield  {journal} {\bibinfo  {journal}
  {Phys.~Rev.~Lett.}\ }\textbf {\bibinfo {volume} {104}},\ \bibinfo {pages}
  {255004} (\bibinfo {year} {2010})}\BibitemShut {NoStop}%
\bibitem [{\citenamefont {Egedal}\ \emph {et~al.}(2011)\citenamefont {Egedal},
  \citenamefont {Katz}, \citenamefont {Bonde}, \citenamefont {Fox},
  \citenamefont {Le}, \citenamefont {Porkolab},\ and\ \citenamefont
  {Vrublevskis}}]{Egedal11}%
  \BibitemOpen
  \bibfield  {author} {\bibinfo {author} {\bibfnamefont {J.}~\bibnamefont
  {Egedal}}, \bibinfo {author} {\bibfnamefont {N.}~\bibnamefont {Katz}},
  \bibinfo {author} {\bibfnamefont {J.}~\bibnamefont {Bonde}}, \bibinfo
  {author} {\bibfnamefont {W.}~\bibnamefont {Fox}}, \bibinfo {author}
  {\bibfnamefont {A.}~\bibnamefont {Le}}, \bibinfo {author} {\bibfnamefont
  {M.}~\bibnamefont {Porkolab}}, \ and\ \bibinfo {author} {\bibfnamefont
  {A.}~\bibnamefont {Vrublevskis}},\ }\bibfield  {title} {\enquote {\bibinfo
  {title} {Spontaneous onset of magnetic reconnection in toroidal plasma caused
  by breaking of 2{D} symmetry},}\ }\href@noop {} {\bibfield  {journal}
  {\bibinfo  {journal} {Phys. Plasmas}\ }\textbf {\bibinfo {volume} {18}},\
  \bibinfo {pages} {111203} (\bibinfo {year} {2011})}\BibitemShut {NoStop}%
\bibitem [{\citenamefont {Dorfman}\ \emph {et~al.}(2013)\citenamefont
  {Dorfman}, \citenamefont {Ji}, \citenamefont {Yamada}, \citenamefont {Yoo},
  \citenamefont {Lawrence}, \citenamefont {Myers},\ and\ \citenamefont
  {Tharp}}]{Dorfman13}%
  \BibitemOpen
  \bibfield  {author} {\bibinfo {author} {\bibfnamefont {S.}~\bibnamefont
  {Dorfman}}, \bibinfo {author} {\bibfnamefont {H.}~\bibnamefont {Ji}},
  \bibinfo {author} {\bibfnamefont {M.}~\bibnamefont {Yamada}}, \bibinfo
  {author} {\bibfnamefont {J.}~\bibnamefont {Yoo}}, \bibinfo {author}
  {\bibfnamefont {E.}~\bibnamefont {Lawrence}}, \bibinfo {author}
  {\bibfnamefont {C.}~\bibnamefont {Myers}}, \ and\ \bibinfo {author}
  {\bibfnamefont {T.~D.}\ \bibnamefont {Tharp}},\ }\bibfield  {title} {\enquote
  {\bibinfo {title} {Three-dimensional, impulsive magnetic reconnection in a
  laboratory plasma},}\ }\href@noop {} {\bibfield  {journal} {\bibinfo
  {journal} {Geophys.~Res.~Lett.}\ }\textbf {\bibinfo {volume} {40}},\ \bibinfo
  {pages} {1} (\bibinfo {year} {2013})}\BibitemShut {NoStop}%
\bibitem [{\citenamefont {Dorfman}\ \emph {et~al.}(2014)\citenamefont
  {Dorfman}, \citenamefont {Ji}, \citenamefont {Yamada}, \citenamefont {Yoo},
  \citenamefont {Lawrence}, \citenamefont {Myers},\ and\ \citenamefont
  {Tharp}}]{Dorfman14}%
  \BibitemOpen
  \bibfield  {author} {\bibinfo {author} {\bibfnamefont {S.}~\bibnamefont
  {Dorfman}}, \bibinfo {author} {\bibfnamefont {H.}~\bibnamefont {Ji}},
  \bibinfo {author} {\bibfnamefont {M.}~\bibnamefont {Yamada}}, \bibinfo
  {author} {\bibfnamefont {J.}~\bibnamefont {Yoo}}, \bibinfo {author}
  {\bibfnamefont {E.}~\bibnamefont {Lawrence}}, \bibinfo {author}
  {\bibfnamefont {C.}~\bibnamefont {Myers}}, \ and\ \bibinfo {author}
  {\bibfnamefont {T.~D.}\ \bibnamefont {Tharp}},\ }\bibfield  {title} {\enquote
  {\bibinfo {title} {Experimental observation of 3-{D}, impulsive reconnection
  events in a laboratory plasma},}\ }\href {\doibase 10.1063/1.4862039}
  {\bibfield  {journal} {\bibinfo  {journal} {Physics of Plasmas}\ }\textbf
  {\bibinfo {volume} {21}},\ \bibinfo {pages} {012109} (\bibinfo {year}
  {2014})}\BibitemShut {NoStop}%
\bibitem [{\citenamefont {Phan}\ \emph {et~al.}(2006)\citenamefont {Phan},
  \citenamefont {Gosling}, \citenamefont {Davis}, \citenamefont {Skoug},
  \citenamefont {Oieroset}, \citenamefont {Lin}, \citenamefont {Lepping},
  \citenamefont {McComas}, \citenamefont {Smith}, \citenamefont {Reme},\ and\
  \citenamefont {Balogh}}]{Phan06}%
  \BibitemOpen
  \bibfield  {author} {\bibinfo {author} {\bibfnamefont {T.~D.}\ \bibnamefont
  {Phan}}, \bibinfo {author} {\bibfnamefont {J.~T.}\ \bibnamefont {Gosling}},
  \bibinfo {author} {\bibfnamefont {M.~S.}\ \bibnamefont {Davis}}, \bibinfo
  {author} {\bibfnamefont {R.~M.}\ \bibnamefont {Skoug}}, \bibinfo {author}
  {\bibfnamefont {M.}~\bibnamefont {Oieroset}}, \bibinfo {author}
  {\bibfnamefont {R.~P.}\ \bibnamefont {Lin}}, \bibinfo {author} {\bibfnamefont
  {R.~P.}\ \bibnamefont {Lepping}}, \bibinfo {author} {\bibfnamefont {D.~J.}\
  \bibnamefont {McComas}}, \bibinfo {author} {\bibfnamefont {C.~W.}\
  \bibnamefont {Smith}}, \bibinfo {author} {\bibfnamefont {H.}~\bibnamefont
  {Reme}}, \ and\ \bibinfo {author} {\bibfnamefont {A.}~\bibnamefont
  {Balogh}},\ }\bibfield  {title} {\enquote {\bibinfo {title} {A magnetic
  reconnection {X}-line extending more than 390 {E}arth radii in the solar
  wind},}\ }\href@noop {} {\bibfield  {journal} {\bibinfo  {journal} {Nature}\
  }\textbf {\bibinfo {volume} {439}},\ \bibinfo {pages} {175} (\bibinfo {year}
  {2006})}\BibitemShut {NoStop}%
\bibitem [{\citenamefont {Gosling}\ \emph {et~al.}(2007)\citenamefont
  {Gosling}, \citenamefont {Eriksson}, \citenamefont {Blush}, \citenamefont
  {Phan}, \citenamefont {Luhmann}, \citenamefont {McComas}, \citenamefont
  {Skoug}, \citenamefont {Acuna}, \citenamefont {Russell},\ and\ \citenamefont
  {Simunac}}]{Gosling07c}%
  \BibitemOpen
  \bibfield  {author} {\bibinfo {author} {\bibfnamefont {J.~T.}\ \bibnamefont
  {Gosling}}, \bibinfo {author} {\bibfnamefont {S.}~\bibnamefont {Eriksson}},
  \bibinfo {author} {\bibfnamefont {L.~M.}\ \bibnamefont {Blush}}, \bibinfo
  {author} {\bibfnamefont {T.~D.}\ \bibnamefont {Phan}}, \bibinfo {author}
  {\bibfnamefont {J.~G.}\ \bibnamefont {Luhmann}}, \bibinfo {author}
  {\bibfnamefont {D.~J.}\ \bibnamefont {McComas}}, \bibinfo {author}
  {\bibfnamefont {R.~M.}\ \bibnamefont {Skoug}}, \bibinfo {author}
  {\bibfnamefont {M.~H.}\ \bibnamefont {Acuna}}, \bibinfo {author}
  {\bibfnamefont {C.~T.}\ \bibnamefont {Russell}}, \ and\ \bibinfo {author}
  {\bibfnamefont {K.~D.}\ \bibnamefont {Simunac}},\ }\bibfield  {title}
  {\enquote {\bibinfo {title} {Five spacecraft observations of oppositely
  directed exhaust jets from a magnetic reconnection {X}-line extending {>}
  4.26 {$\times 10^6$} km in the solar wind at 1 {AU}},}\ }\href@noop {}
  {\bibfield  {journal} {\bibinfo  {journal} {Geophys.~Res.~Lett.}\ }\textbf
  {\bibinfo {volume} {34}},\ \bibinfo {pages} {L20108} (\bibinfo {year}
  {2007})}\BibitemShut {NoStop}%
\bibitem [{\citenamefont {Huba}\ and\ \citenamefont {Rudakov}(2002)}]{huba02}%
  \BibitemOpen
  \bibfield  {author} {\bibinfo {author} {\bibfnamefont {J.~D.}\ \bibnamefont
  {Huba}}\ and\ \bibinfo {author} {\bibfnamefont {L.~I.}\ \bibnamefont
  {Rudakov}},\ }\bibfield  {title} {\enquote {\bibinfo {title}
  {Three-dimensional {H}all magnetic reconnection},}\ }\href@noop {} {\bibfield
   {journal} {\bibinfo  {journal} {Phys. Plasmas}\ }\textbf {\bibinfo {volume}
  {9}},\ \bibinfo {pages} {4435} (\bibinfo {year} {2002})}\BibitemShut
  {NoStop}%
\bibitem [{\citenamefont {Huba}\ and\ \citenamefont {Rudakov}(2003)}]{Huba03}%
  \BibitemOpen
  \bibfield  {author} {\bibinfo {author} {\bibfnamefont {J.~D.}\ \bibnamefont
  {Huba}}\ and\ \bibinfo {author} {\bibfnamefont {L.~I.}\ \bibnamefont
  {Rudakov}},\ }\bibfield  {title} {\enquote {\bibinfo {title} {Hall
  magnetohydrodynamics of neutral layers},}\ }\href@noop {} {\bibfield
  {journal} {\bibinfo  {journal} {Phys.~Plasmas}\ }\textbf {\bibinfo {volume}
  {10}},\ \bibinfo {pages} {3139} (\bibinfo {year} {2003})}\BibitemShut
  {NoStop}%
\bibitem [{\citenamefont {Karimabadi}\ \emph {et~al.}(2004)\citenamefont
  {Karimabadi}, \citenamefont {Krauss-Varban}, \citenamefont {Huba},\ and\
  \citenamefont {Vu}}]{Karimabadi04}%
  \BibitemOpen
  \bibfield  {author} {\bibinfo {author} {\bibfnamefont {H.}~\bibnamefont
  {Karimabadi}}, \bibinfo {author} {\bibfnamefont {D.}~\bibnamefont
  {Krauss-Varban}}, \bibinfo {author} {\bibfnamefont {J.~D.}\ \bibnamefont
  {Huba}}, \ and\ \bibinfo {author} {\bibfnamefont {H.~X.}\ \bibnamefont
  {Vu}},\ }\bibfield  {title} {\enquote {\bibinfo {title} {On magnetic
  reconnection regimes and associated three-dimensional asymmetries: {H}ybrid,
  {H}all-less hybrid, and {H}all-{MHD} simulations},}\ }\href@noop {}
  {\bibfield  {journal} {\bibinfo  {journal} {J.~Geophys.~Res.}\ }\textbf
  {\bibinfo {volume} {109}},\ \bibinfo {pages} {A09205} (\bibinfo {year}
  {2004})}\BibitemShut {NoStop}%
\bibitem [{\citenamefont {Lapenta}\ \emph {et~al.}(2006)\citenamefont
  {Lapenta}, \citenamefont {Krauss-Varban}, \citenamefont {Karimabadi},
  \citenamefont {Huba}, \citenamefont {Rudakov},\ and\ \citenamefont
  {Ricci}}]{Lapenta06}%
  \BibitemOpen
  \bibfield  {author} {\bibinfo {author} {\bibfnamefont {G.}~\bibnamefont
  {Lapenta}}, \bibinfo {author} {\bibfnamefont {D.}~\bibnamefont
  {Krauss-Varban}}, \bibinfo {author} {\bibfnamefont {H.}~\bibnamefont
  {Karimabadi}}, \bibinfo {author} {\bibfnamefont {J.~D.}\ \bibnamefont
  {Huba}}, \bibinfo {author} {\bibfnamefont {L.~I.}\ \bibnamefont {Rudakov}}, \
  and\ \bibinfo {author} {\bibfnamefont {P.}~\bibnamefont {Ricci}},\ }\bibfield
   {title} {\enquote {\bibinfo {title} {Kinetic simulations of x-line expansion
  in 3{D} reconnection},}\ }\href@noop {} {\bibfield  {journal} {\bibinfo
  {journal} {Geophys.~Res.~Lett.}\ }\textbf {\bibinfo {volume} {33}},\ \bibinfo
  {pages} {L10102} (\bibinfo {year} {2006})}\BibitemShut {NoStop}%
\bibitem [{\citenamefont {Shepherd}\ and\ \citenamefont
  {Cassak}(2012)}]{Shepherd12}%
  \BibitemOpen
  \bibfield  {author} {\bibinfo {author} {\bibfnamefont {L.~S.}\ \bibnamefont
  {Shepherd}}\ and\ \bibinfo {author} {\bibfnamefont {P.~A.}\ \bibnamefont
  {Cassak}},\ }\bibfield  {title} {\enquote {\bibinfo {title} {Guide field
  dependence of 3{D} {X}-line spreading during collisionless magnetic
  reconnection},}\ }\href@noop {} {\bibfield  {journal} {\bibinfo  {journal}
  {J.~Geophys.~Res.}\ }\textbf {\bibinfo {volume} {117}},\ \bibinfo {pages}
  {A10101} (\bibinfo {year} {2012})}\BibitemShut {NoStop}%
\bibitem [{\citenamefont {Nakamura}\ \emph {et~al.}(2012)\citenamefont
  {Nakamura}, \citenamefont {Nakamura}, \citenamefont {Alexandrova},
  \citenamefont {Kubota},\ and\ \citenamefont {Nagai}}]{Nakamura12}%
  \BibitemOpen
  \bibfield  {author} {\bibinfo {author} {\bibfnamefont {T.~K.~M.}\
  \bibnamefont {Nakamura}}, \bibinfo {author} {\bibfnamefont {R.}~\bibnamefont
  {Nakamura}}, \bibinfo {author} {\bibfnamefont {A.}~\bibnamefont
  {Alexandrova}}, \bibinfo {author} {\bibfnamefont {Y.}~\bibnamefont {Kubota}},
  \ and\ \bibinfo {author} {\bibfnamefont {T.}~\bibnamefont {Nagai}},\
  }\bibfield  {title} {\enquote {\bibinfo {title} {Hall magnetohydrodynamic
  effects for three-dimensional magnetic reconnection with finite width along
  the direction of the current},}\ }\href@noop {} {\bibfield  {journal}
  {\bibinfo  {journal} {J. Geophys. Res.}\ }\textbf {\bibinfo {volume} {117}},\
  \bibinfo {pages} {03220} (\bibinfo {year} {2012})}\BibitemShut {NoStop}%
\bibitem [{\citenamefont {Jain}\ \emph {et~al.}(2013)\citenamefont {Jain},
  \citenamefont {B{\"u}chner}, \citenamefont {Dorfman}, \citenamefont {Ji},\
  and\ \citenamefont {Sharma}}]{Jain13}%
  \BibitemOpen
  \bibfield  {author} {\bibinfo {author} {\bibfnamefont {N.}~\bibnamefont
  {Jain}}, \bibinfo {author} {\bibfnamefont {J.}~\bibnamefont {B{\"u}chner}},
  \bibinfo {author} {\bibfnamefont {S.}~\bibnamefont {Dorfman}}, \bibinfo
  {author} {\bibfnamefont {H.}~\bibnamefont {Ji}}, \ and\ \bibinfo {author}
  {\bibfnamefont {A.~S.}\ \bibnamefont {Sharma}},\ }\bibfield  {title}
  {\enquote {\bibinfo {title} {Current disruption and its spreading in
  collisionless magnetic reconnection},}\ }\href@noop {} {\bibfield  {journal}
  {\bibinfo  {journal} {Phys.~Plasmas}\ }\textbf {\bibinfo {volume} {20}},\
  \bibinfo {pages} {112101} (\bibinfo {year} {2013})}\BibitemShut {NoStop}%
\bibitem [{\citenamefont {Hesse}, \citenamefont {Kuznetsova},\ and\
  \citenamefont {Birn}(2001)}]{Hesse01}%
  \BibitemOpen
  \bibfield  {author} {\bibinfo {author} {\bibfnamefont {M.}~\bibnamefont
  {Hesse}}, \bibinfo {author} {\bibfnamefont {M.}~\bibnamefont {Kuznetsova}}, \
  and\ \bibinfo {author} {\bibfnamefont {J.}~\bibnamefont {Birn}},\ }\bibfield
  {title} {\enquote {\bibinfo {title} {Particle-in-cell simulations of
  three-dimensional collisionless magnetic reconnection},}\ }\href@noop {}
  {\bibfield  {journal} {\bibinfo  {journal} {J. Geophys. Res.}\ }\textbf
  {\bibinfo {volume} {106}},\ \bibinfo {pages} {29831--29841} (\bibinfo {year}
  {2001})}\BibitemShut {NoStop}%
\bibitem [{\citenamefont {Li}\ \emph {et~al.}(2020)\citenamefont {Li},
  \citenamefont {Liu}, \citenamefont {Hesse},\ and\ \citenamefont {Zou}}]{tak}%
  \BibitemOpen
  \bibfield  {author} {\bibinfo {author} {\bibfnamefont {T.}~\bibnamefont
  {Li}}, \bibinfo {author} {\bibfnamefont {Y.-H.}\ \bibnamefont {Liu}},
  \bibinfo {author} {\bibfnamefont {M.}~\bibnamefont {Hesse}}, \ and\ \bibinfo
  {author} {\bibfnamefont {Y.}~\bibnamefont {Zou}},\ }\bibfield  {title}
  {\enquote {\bibinfo {title} {Three-dimensional x-line spreading in asymmetric
  magnetic reconnection},}\ }\href {\doibase 10.1029/2019JA027094} {\bibfield
  {journal} {\bibinfo  {journal} {Journal of Geophysical Research: Space
  Physics}\ }\textbf {\bibinfo {volume} {125}},\ \bibinfo {pages}
  {e2019JA027094} (\bibinfo {year} {2020})}\BibitemShut {NoStop}%
\bibitem [{\citenamefont {Gosling}\ \emph {et~al.}(2005)\citenamefont
  {Gosling}, \citenamefont {Skoug}, \citenamefont {McComas},\ and\
  \citenamefont {Smith}}]{Gosling05a}%
  \BibitemOpen
  \bibfield  {author} {\bibinfo {author} {\bibfnamefont {J.~T.}\ \bibnamefont
  {Gosling}}, \bibinfo {author} {\bibfnamefont {R.~M.}\ \bibnamefont {Skoug}},
  \bibinfo {author} {\bibfnamefont {D.~J.}\ \bibnamefont {McComas}}, \ and\
  \bibinfo {author} {\bibfnamefont {C.~W.}\ \bibnamefont {Smith}},\ }\bibfield
  {title} {\enquote {\bibinfo {title} {Direct evidence for magnetic
  reconnection in the solar wind near 1 {AU}},}\ }\href {\doibase
  10.1029/2004JA010809} {\bibfield  {journal} {\bibinfo  {journal} {J. Geophys.
  Res.}\ }\textbf {\bibinfo {volume} {110}},\ \bibinfo {eid} {A01107} (\bibinfo
  {year} {2005}),\ 10.1029/2004JA010809}\BibitemShut {NoStop}%
\bibitem [{\citenamefont {Jain}\ and\ \citenamefont
  {B{\"u}chner}(2017)}]{Jain17}%
  \BibitemOpen
  \bibfield  {author} {\bibinfo {author} {\bibfnamefont {N.}~\bibnamefont
  {Jain}}\ and\ \bibinfo {author} {\bibfnamefont {J.}~\bibnamefont
  {B{\"u}chner}},\ }\bibfield  {title} {\enquote {\bibinfo {title} {Spreading
  of electron scale magnetic reconnection with a wave number dependent speed
  due to the propagation of dispersive waves},}\ }\href {\doibase
  10.1063/1.4994704} {\bibfield  {journal} {\bibinfo  {journal} {Physics of
  Plasmas}\ }\textbf {\bibinfo {volume} {24}},\ \bibinfo {pages} {082304}
  (\bibinfo {year} {2017})}\BibitemShut {NoStop}%
\bibitem [{\citenamefont {Schreier}\ \emph {et~al.}(2010)\citenamefont
  {Schreier}, \citenamefont {Swisdak}, \citenamefont {Drake},\ and\
  \citenamefont {Cassak}}]{Schreier10}%
  \BibitemOpen
  \bibfield  {author} {\bibinfo {author} {\bibfnamefont {R.}~\bibnamefont
  {Schreier}}, \bibinfo {author} {\bibfnamefont {M.}~\bibnamefont {Swisdak}},
  \bibinfo {author} {\bibfnamefont {J.~F.}\ \bibnamefont {Drake}}, \ and\
  \bibinfo {author} {\bibfnamefont {P.~A.}\ \bibnamefont {Cassak}},\ }\bibfield
   {title} {\enquote {\bibinfo {title} {Three-dimensional simulations of the
  orientation and structure of reconnection {X}-lines},}\ }\href@noop {}
  {\bibfield  {journal} {\bibinfo  {journal} {Phys.~Plasmas}\ }\textbf
  {\bibinfo {volume} {17}},\ \bibinfo {pages} {110704} (\bibinfo {year}
  {2010})}\BibitemShut {NoStop}%
\bibitem [{\citenamefont {Vorpahl}(1976)}]{Vorpahl76}%
  \BibitemOpen
  \bibfield  {author} {\bibinfo {author} {\bibfnamefont {J.~A.}\ \bibnamefont
  {Vorpahl}},\ }\bibfield  {title} {\enquote {\bibinfo {title} {The triggering
  and subsequent development of a solar flare},}\ }\href {\doibase
  10.1086/154343} {\bibfield  {journal} {\bibinfo  {journal} {ApJ}\ }\textbf
  {\bibinfo {volume} {205}},\ \bibinfo {pages} {868} (\bibinfo {year}
  {1976})}\BibitemShut {NoStop}%
\bibitem [{\citenamefont {Shay}\ \emph {et~al.}(2004)\citenamefont {Shay},
  \citenamefont {Drake}, \citenamefont {Swisdak},\ and\ \citenamefont
  {Rogers}}]{Shay04}%
  \BibitemOpen
  \bibfield  {author} {\bibinfo {author} {\bibfnamefont {M.~A.}\ \bibnamefont
  {Shay}}, \bibinfo {author} {\bibfnamefont {J.~F.}\ \bibnamefont {Drake}},
  \bibinfo {author} {\bibfnamefont {M.}~\bibnamefont {Swisdak}}, \ and\
  \bibinfo {author} {\bibfnamefont {B.~N.}\ \bibnamefont {Rogers}},\ }\bibfield
   {title} {\enquote {\bibinfo {title} {The scaling of embedded collisionless
  reconnection},}\ }\href@noop {} {\bibfield  {journal} {\bibinfo  {journal}
  {Phys. Plasmas}\ }\textbf {\bibinfo {volume} {11}},\ \bibinfo {pages} {2199}
  (\bibinfo {year} {2004})}\BibitemShut {NoStop}%
\bibitem [{\citenamefont {Guzdar}\ \emph {et~al.}(1993)\citenamefont {Guzdar},
  \citenamefont {Drake}, \citenamefont {McCarthy}, \citenamefont {Hassam},\
  and\ \citenamefont {Liu}}]{Guzdar93}%
  \BibitemOpen
  \bibfield  {author} {\bibinfo {author} {\bibfnamefont {P.~N.}\ \bibnamefont
  {Guzdar}}, \bibinfo {author} {\bibfnamefont {J.~F.}\ \bibnamefont {Drake}},
  \bibinfo {author} {\bibfnamefont {D.}~\bibnamefont {McCarthy}}, \bibinfo
  {author} {\bibfnamefont {A.~B.}\ \bibnamefont {Hassam}}, \ and\ \bibinfo
  {author} {\bibfnamefont {C.~S.}\ \bibnamefont {Liu}},\ }\bibfield  {title}
  {\enquote {\bibinfo {title} {Three-dimensional fluid simulations of the
  nonlinear drift-resistive ballooning modes in tokamak edge plasmas},}\
  }\href@noop {} {\bibfield  {journal} {\bibinfo  {journal} {Phys.~Fluids B}\
  }\textbf {\bibinfo {volume} {5}},\ \bibinfo {pages} {3712--3727} (\bibinfo
  {year} {1993})}\BibitemShut {NoStop}%
\bibitem [{\citenamefont {Shay}\ \emph {et~al.}(1999)\citenamefont {Shay},
  \citenamefont {Drake}, \citenamefont {Rogers},\ and\ \citenamefont
  {Denton}}]{shay99a}%
  \BibitemOpen
  \bibfield  {author} {\bibinfo {author} {\bibfnamefont {M.~A.}\ \bibnamefont
  {Shay}}, \bibinfo {author} {\bibfnamefont {J.~F.}\ \bibnamefont {Drake}},
  \bibinfo {author} {\bibfnamefont {B.~N.}\ \bibnamefont {Rogers}}, \ and\
  \bibinfo {author} {\bibfnamefont {R.~E.}\ \bibnamefont {Denton}},\ }\bibfield
   {title} {\enquote {\bibinfo {title} {The scaling of collisionless, magnetic
  reconnection for large systems},}\ }\href@noop {} {\bibfield  {journal}
  {\bibinfo  {journal} {Geophys. Res. Lett.}\ }\textbf {\bibinfo {volume}
  {26}},\ \bibinfo {pages} {2163--2166} (\bibinfo {year} {1999})}\BibitemShut
  {NoStop}%
\bibitem [{\citenamefont {Birn}\ \emph {et~al.}(2001)\citenamefont {Birn},
  \citenamefont {Drake}, \citenamefont {Shay}, \citenamefont {Rogers},
  \citenamefont {Denton}, \citenamefont {Hesse}, \citenamefont {Kuznetsova},
  \citenamefont {Ma}, \citenamefont {Bhattacharjee}, \citenamefont {Otto},\
  and\ \citenamefont {Pritchett}}]{Birn01}%
  \BibitemOpen
  \bibfield  {author} {\bibinfo {author} {\bibfnamefont {J.}~\bibnamefont
  {Birn}}, \bibinfo {author} {\bibfnamefont {J.~F.}\ \bibnamefont {Drake}},
  \bibinfo {author} {\bibfnamefont {M.~A.}\ \bibnamefont {Shay}}, \bibinfo
  {author} {\bibfnamefont {B.~N.}\ \bibnamefont {Rogers}}, \bibinfo {author}
  {\bibfnamefont {R.~E.}\ \bibnamefont {Denton}}, \bibinfo {author}
  {\bibfnamefont {M.}~\bibnamefont {Hesse}}, \bibinfo {author} {\bibfnamefont
  {M.}~\bibnamefont {Kuznetsova}}, \bibinfo {author} {\bibfnamefont {Z.~W.}\
  \bibnamefont {Ma}}, \bibinfo {author} {\bibfnamefont {A.}~\bibnamefont
  {Bhattacharjee}}, \bibinfo {author} {\bibfnamefont {A.}~\bibnamefont {Otto}},
  \ and\ \bibinfo {author} {\bibfnamefont {P.~L.}\ \bibnamefont {Pritchett}},\
  }\bibfield  {title} {\enquote {\bibinfo {title} {{GEM} magnetic reconnection
  challenge},}\ }\href@noop {} {\bibfield  {journal} {\bibinfo  {journal} {J.
  Geophys. Res.}\ }\textbf {\bibinfo {volume} {106}},\ \bibinfo {pages} {3715}
  (\bibinfo {year} {2001})}\BibitemShut {NoStop}%
\bibitem [{\citenamefont {Liu}\ \emph {et~al.}(2017)\citenamefont {Liu},
  \citenamefont {Hesse}, \citenamefont {Guo}, \citenamefont {Daughton},
  \citenamefont {Li}, \citenamefont {Cassak},\ and\ \citenamefont
  {Shay}}]{Liu17}%
  \BibitemOpen
  \bibfield  {author} {\bibinfo {author} {\bibfnamefont {Y.-H.}\ \bibnamefont
  {Liu}}, \bibinfo {author} {\bibfnamefont {M.}~\bibnamefont {Hesse}}, \bibinfo
  {author} {\bibfnamefont {F.}~\bibnamefont {Guo}}, \bibinfo {author}
  {\bibfnamefont {W.}~\bibnamefont {Daughton}}, \bibinfo {author}
  {\bibfnamefont {H.}~\bibnamefont {Li}}, \bibinfo {author} {\bibfnamefont
  {P.~A.}\ \bibnamefont {Cassak}}, \ and\ \bibinfo {author} {\bibfnamefont
  {M.~A.}\ \bibnamefont {Shay}},\ }\bibfield  {title} {\enquote {\bibinfo
  {title} {Why does steady-state magnetic reconnection have a maximum local
  rate of order 0.1?}}\ }\href@noop {} {\bibfield  {journal} {\bibinfo
  {journal} {Phys.~Rev.~Lett.}\ }\textbf {\bibinfo {volume} {118}},\ \bibinfo
  {pages} {085101} (\bibinfo {year} {2017})}\BibitemShut {NoStop}%
\bibitem [{\citenamefont {Ugai}\ and\ \citenamefont {Tsuda}(1977)}]{Ugai77}%
  \BibitemOpen
  \bibfield  {author} {\bibinfo {author} {\bibfnamefont {M.}~\bibnamefont
  {Ugai}}\ and\ \bibinfo {author} {\bibfnamefont {T.}~\bibnamefont {Tsuda}},\
  }\bibfield  {title} {\enquote {\bibinfo {title} {Magnetic field line
  reconnexion by localized enhancement of resistivity, 1, evolution in a
  compressible mhd fluid},}\ }\href@noop {} {\bibfield  {journal} {\bibinfo
  {journal} {J.~Plasma Phys.}\ }\textbf {\bibinfo {volume} {17}},\ \bibinfo
  {pages} {337} (\bibinfo {year} {1977})}\BibitemShut {NoStop}%
\bibitem [{\citenamefont {Sato}\ and\ \citenamefont {Hayashi}(1979)}]{Sato79}%
  \BibitemOpen
  \bibfield  {author} {\bibinfo {author} {\bibfnamefont {T.}~\bibnamefont
  {Sato}}\ and\ \bibinfo {author} {\bibfnamefont {T.}~\bibnamefont {Hayashi}},\
  }\bibfield  {title} {\enquote {\bibinfo {title} {Externally driven magnetic
  reconnection and a powerful magnetic energy converter},}\ }\href@noop {}
  {\bibfield  {journal} {\bibinfo  {journal} {Phys. Fluids}\ }\textbf {\bibinfo
  {volume} {22}},\ \bibinfo {pages} {1189} (\bibinfo {year}
  {1979})}\BibitemShut {NoStop}%
\end{thebibliography}

%\appendix
%\section{appendix section}

\end{document}